\pdfoutput=1
\documentclass[cernpreprint, atlasdraft=false, texlive=2016, UKenglish, texmf]{atlasdoc}
 
\usepackage[block=none]{atlaspackage}
\usepackage{atlasbiblatex}
 
\usepackage{atlasphysics}
 
\graphicspath{{logos/}{figures/}}
 
\usepackage{ANA-BPHY-2018-08-PAPER-defs}

\AtlasTitle{Study of $B_c^+\to J/\psi D_s^+$ and $B_c^+\to J/\psi D_s^{*+}$ decays
in $pp$ collisions at $\sqrt{s} = 13$\,\TeV\ with the ATLAS detector}
 
\AtlasAbstract{
A study of $B_c^+\to J/\psi D_s^+$ and $B_c^+\to J/\psi D_s^{*+}$ decays
using 139\,fb$^{-1}$ of integrated luminosity collected with the ATLAS detector
from $\sqrt{s} = 13$\,\TeV\ $pp$ collisions at the LHC is presented.
The ratios of the branching fractions of the two decays to the
branching fraction of the $B_c^+\to J/\psi \pi^+$ decay are measured:
$\mathcal B(B_c^+\to J/\psi D_s^+)/\mathcal B(B_c^+\to J/\psi \pi^+) = 2.76\pm 0.47$ and
$\mathcal B(B_c^+\to J/\psi D_s^{*+})/\mathcal B(B_c^+\to J/\psi \pi^+) = 5.33\pm 0.96$.
The ratio of the branching fractions of the two decays is found to be
$\mathcal B(B_c^+\to J/\psi D_s^{*+})/\mathcal B(B_c^+\to J/\psi D_s^+) = 1.93\pm0.26$.
For the $B_c^+\to J/\psi D_s^{*+}$ decay, the transverse polarization fraction, $\Gamma_{\pm\pm}/\Gamma$, is measured
to be $0.70\pm0.11$.
The reported uncertainties include both the statistical and systematic components added in quadrature.
The precision of the measurements exceeds that in
all previous studies of these decays.
These results supersede those obtained in the earlier ATLAS study of the same decays
with $\sqrt{s} = 7$ and 8\,\TeV\ $pp$ collision data.
A comparison with available theoretical predictions for the measured quantities is presented.
}
 
\author{The ATLAS Collaboration}
 
\AtlasRefCode{BPHY-2018-08}

\AtlasJournal{JHEP}
 
\PreprintIdNumber{CERN-EP-2022-025}
 
\AtlasJournalRef{JHEP 08 (2022) 087}
 
\AtlasDOI{10.1007/JHEP08(2022)087}
\hypersetup{pdftitle={ATLAS document},pdfauthor={The ATLAS Collaboration}}
 
\begin{document}
 
\maketitle

\section{Introduction}
\label{sec:intro}
The large amount of \pp collision data collected by the LHC experiments at a centre-of-mass energy $\sqs = 13$\,\TeV\
between 2015 and 2018 (\RunTwo) opens
new opportunities to measure the properties of \Bcp mesons precisely. Previous studies were limited
by the low \Bcp production cross-section.
Being the only weakly decaying meson consisting of two heavy
quarks, the \Bcp meson provides a unique testing ground for various theoretical approaches
that are used to describe its production and decays.
In particular, the \Bcp decays can occur through a weak transition of either heavy quark
as well as through a weak annihilation.

\begin{figure*}[htbp]
\centering
\subfloat[]{
\includegraphics[width=0.3\textwidth]{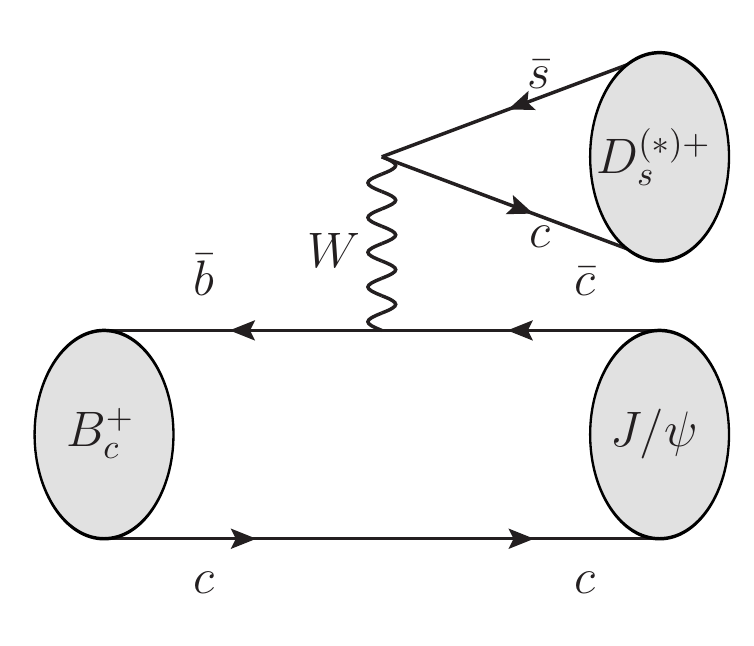}
\label{fig:Diagrams:Diag_spectator}
}
\subfloat[]{
\includegraphics[width=0.3\textwidth]{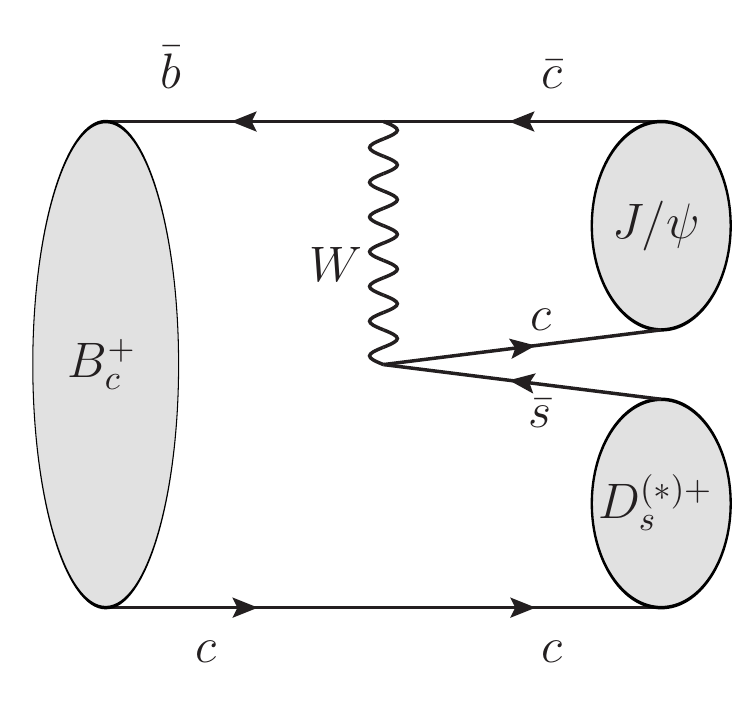}
\label{fig:Diagrams:Diag_spectator_colsup}
}
\subfloat[]{
\includegraphics[width=0.3\textwidth]{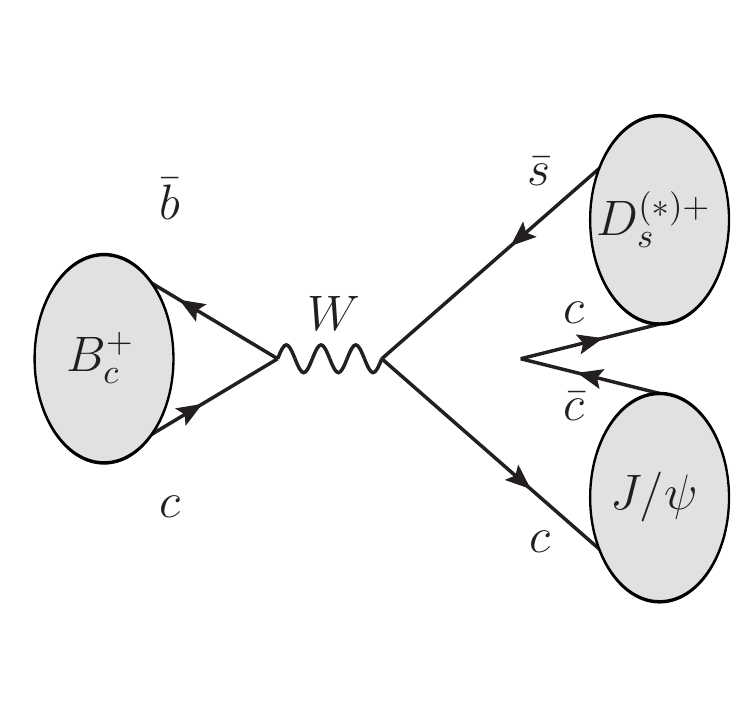}
\label{fig:Diagrams:Diag_annihil}
}
\caption{Feynman diagrams for $\BcJpsiDsOrStar$ decays:
\protect\subref{fig:Diagrams:Diag_spectator} colour-favoured spectator,
\protect\subref{fig:Diagrams:Diag_spectator_colsup} colour-suppressed
spectator, and
\protect\subref{fig:Diagrams:Diag_annihil} annihilation topology.
\label{fig:Diagrams}}
\end{figure*}

Decays of \Bcp to final states with a \Jpsi meson involve a $\bar b$-quark transition, with the
$c$-quark being a spectator. There is also a contribution from an annihilation diagram.
Examples of the corresponding diagrams for the \BcJpsiDs and \BcJpsiDsStar decays are shown in Figure~\ref{fig:Diagrams}.
 
These decays were first observed by the LHCb experiment~\cite{LHCb:2013kwl} and later by ATLAS~\cite{BPHY-2014-04} using
data collected at centre-of-mass energies $\sqs = 7$ and 8~\TeV\ in 2011--2012 (\RunOne).
Various predictions of their properties exist in
theoretical works~\cite{Colangelo:1999zn, Kiselev:2002vz, Ivanov:2006ni, Dubnicka:2017job,
Dhir:2008hh, Ke:2013yka, Rui:2014tpa, Kar:2013fna, Nayak:2022qaq, Mohammadi:2018jcp}.
Only calculations in Refs.~\cite{Rui:2014tpa,Mohammadi:2018jcp} explicitly account
for small contributions from the annihilation diagram (Figure~\ref{fig:Diagrams:Diag_annihil})
to the \BcJpsiDsOrStar decay amplitudes.
 
This paper presents a new study of these decays with the ATLAS detector~\cite{PERF-2007-01}.
It is based on \pp collision data
collected at $\sqs = 13$\,\TeV\ during \RunTwo\ of the LHC,
corresponding to an integrated luminosity of
139\,\ifb. The purpose of the new measurement is to improve the precision of measured properties of these decays
by using a larger
data sample and new analysis methods. In particular, a technique using boosted decision trees is applied
to improve the signal decay candidate selection.
 
Similarly to Ref.~\cite{BPHY-2014-04}, the \Ds meson is reconstructed via the $\Ds\to\phi\pip$ decay,
with the $\phi$ meson decaying into a pair of charged kaons. The \DsStar meson decays
into a \Ds meson and a soft photon or $\pi^0$ which is not reconstructed in
the analysis. However, the mass difference between the \DsStar and \Ds mesons is sufficient for
the two decay signals to be resolved as two distinct structures in the distribution of the
reconstructed mass of the $\Jpsi\Ds$ system.
The \Jpsi meson is reconstructed via its decay into a muon pair.
 
The \BcJpsiPi decay is used as a reference to measure the branching fractions. The following
ratios are measured: $\RDsPi = \Br(\BcJpsiDs)/\Br(\BcJpsiPi)$, $\RDsStarPi = \Br(\BcJpsiDsStar)/\Br(\BcJpsiPi)$,
and $\RDsStarDs = \Br(\BcJpsiDsStar)/\Br(\BcJpsiDs)$.
 
The \BcJpsiDsStar decay, being a transition of a pseudoscalar meson into two vector states,
can be described in terms of three helicity amplitudes: \APP,
\Azz and \AMM, where the indices correspond to the helicities
of the \Jpsi and \DsStar mesons.  The contribution of the \APP and \AMM amplitudes,
referred to as the \App component, corresponds to the \Jpsi and \DsStar
transverse polarization.
Its fraction, \Gpp, is also measured.

\section{The ATLAS detector, data and simulated samples}
\label{sec:detector}
The ATLAS detector\footnote{
ATLAS uses a right-handed coordinate system with its origin at the nominal interaction point (IP)
in the centre of the detector and the \(z\)-axis along the beam pipe.
The \(x\)-axis points from the IP to the centre of the LHC ring,
and the \(y\)-axis points upward.
Cylindrical coordinates \((r,\phi)\) are used in the transverse plane,
\(\phi\) being the azimuthal angle around the \(z\)-axis and \(r\) the distance from the IP in the transverse plane.
The pseudorapidity is defined in terms of the polar angle \(\theta\) as \(\eta = -\ln \tan(\theta/2)\).
Angular distance is measured in units of \(\Delta R \equiv \sqrt{(\Delta\eta)^{2} + (\Delta\phi)^{2}}\).
}
consists of three main components: an inner detector (ID) tracking system immersed
in a 2~T axial magnetic field, surrounded by electromagnetic and hadronic calorimeters and by the muon
spectrometer (MS).
A full description of the detector
can be found in Ref.~\cite{PERF-2007-01}, complemented by
Ref.~\cite{PIX-2018-001} for details
about the innermost silicon pixel layer that was installed for \RunTwo.
 
This analysis is based on the full \RunTwo dataset of $\sqs = 13$\,\TeV\ \pp collisions collected by ATLAS between 2015 and 2018
at the LHC. The data were recorded during stable LHC beam periods.
Data quality requirements were imposed, notably on the performance of the MS and ID systems~\cite{DAPR-2018-01}.
After applying these criteria, the integrated luminosity of the dataset amounts to 139\,\ifb, with an uncertainty
of 1.7\%~\cite{ATLAS-CONF-2019-021}, obtained using the LUCID-2 detector~\cite{Avoni:2018iuv}
for the primary luminosity measurements.
 
The data were collected during periods with different instantaneous luminosities, so several trigger configurations were
used in the analysis to maximize the signal yields.
Most of the triggers were based on identification of two muons,
with various muon transverse momentum (\pT) thresholds (usually 4 and 6\,\GeV),
requiring the oppositely charged muons in the pair to form a good vertex and have an invariant mass compatible with being
produced by a $\Jpsi\to\mumu$ decay.
Since the event rate from these triggers was too high, prescale
factors were applied to reduce it during periods with high instantaneous luminosity.
To allow lower thresholds while keeping acceptable rates, other triggers required
the presence of a third muon,
which may appear in a semileptonic decay of hadrons containing $b$ and $\bar c$ quarks present in an event
along with the \Bcp meson.
Another set of triggers fitted a common vertex to two muons and two additional ID tracks, applying an
invariant mass selection that assumed the $B^0_s\to\mumu\phi$ decay topology.
The events selected by these triggers require a special
treatment described in Section~\ref{sec:fit}.
 
To model the inelastic \pp events containing \Bcp decays, large samples of Monte Carlo (MC) simulated events
were prepared using the dedicated generator BCVEGPY~2.2~\cite{Chang:2015qea}
interfaced to \Pythia[8.244]~\cite{Sjostrand:2014zea} to simulate
the parton showers and hadronization, and \EVTGEN~\cite{Lange:2001uf} to model heavy-flavour decays.
Simulated distributions of $\pT(\Bcp)$ and $|\eta(\Bcp)|$ were corrected to match the corresponding spectra of
\Bcp mesons in data, following the procedure described in Section~\ref{sec:syst}.
The generated events with at least two muons having transverse momenta $\pT(\mu) > 3.5$\,GeV and
pseudorapidities $|\eta(\mu)| < 2.8$ were passed through a full simulation of the detector using the ATLAS
simulation~\cite{SOFT-2010-01} based on \GEANT~\cite{Agostinelli:2002hh}.
A simulation of the triggers and their prescale factors was applied to all
MC event samples used in the analysis.
 
An extensive software suite~\cite{ATL-SOFT-PUB-2021-001} is used in the reconstruction
and analysis of real and simulated data, in detector operations, and in the trigger and data acquisition systems of the experiment.

\section{Reconstruction and candidate selection}
\label{sec:reco}
 
\subsection{\Jpsi candidates}
\label{sec:reco:Jpsi}
The $\Jpsi$ candidates are built from pairs of oppositely charged muon
candidates that are reconstructed using information from the MS and the ID.
Muon candidates must satisfy the \emph{Loose} identification working point~\cite{MUON-2018-03}.
The muon track parameters
are determined from the ID measurement alone, since the precision of the measured track
parameters is dominated by the ID track reconstruction in the $\pt$ range of interest
for this analysis.
Pairs of oppositely charged muon tracks are re-fitted to a common vertex.
The quality of the vertex fit is required to satisfy $\chi^2 < 10$, where $\chi^2$
is the fit quality (the number of degrees of freedom $\ndf = 1$).
This loose selection requirement is chosen to minimize biasing the $\chi^2$ distribution of further cascade fit.
The candidates in the dimuon invariant mass window $2800\,\MeV < m(\mu^+\mu^-)<3400\,\MeV$ are retained
for further analysis.

\subsection{\BcJpsiDsOrStar candidates}
\label{sec:reco:BcJpsiDs}
 
Candidate $\Ds$ mesons from the \BcJpsiDsOrStar decays\footnote{
The analysis considers both \Bcp and $B_c^-$ meson decays.
For simplicity, charge conjugated states are implied from here onwards.
}
are reconstructed
using the decay $\Ds\to\phi\pip$ with $\phi\to K^+K^-$.
Oppositely charged particle tracks
are assigned the kaon mass
and combined in pairs to form $\phi$ candidates.
Any additional track is assigned the pion mass
and combined with the $\phi$ candidate to form a $\Ds$ candidate.
Only three-track combinations successfully fitted to a common vertex with
$\chi^2/\ndf<5$ ($\ndf = 3$) are accepted for further analysis.
The invariant mass of the $\phi$ candidate, $m(K^+K^-)$, is required to be within a $\pm7~\MeV$ range
around the world average $\phi$ mass, $m_\phi=1019.461~\MeV$~\cite{pdg2020}.
The $\Ds$ candidate must have an invariant mass, $m(K^+K^-\pi^+)$, between 1930~\MeV\ and 2010~\MeV,
which spans about $\pm2.5$ standard deviations of the detector's mass resolution.
 
The \BcJpsiDsOrStar candidates are built by combining the selected \Jpsi and \Ds candidates.
The \Jpsi meson decays instantly at the same point as the \Bcp meson does,
whereas the \Ds meson lives long enough to form another displaced vertex.
Following this cascade topology, a two-vertex fit is performed
and the \Ds candidate's momentum is required to point back to the \Bcp vertex~\cite{Kostyukhin:2003daa}.
To improve the mass resolution, the invariant masses of the \Jpsi and \Ds candidates are constrained
to the world average measured masses of the \Jpsi and \Ds mesons~\cite{pdg2020}, respectively.
 
In what follows, the \Bcp invariant mass,
$\pT(\Bcp)$ and $\eta(\Bcp)$ are calculated using parameters of the tracks from
the \Bcp cascade vertex fit.
The \Bcp candidates are required to have $\pT(\Bcp)>15\,\GeV$ and $|\eta(\Bcp)|<2.0$.
The re-fitted muon tracks must have $\pT>4\,\GeV$ and $|\eta|<2.3$,
while the re-fitted hadron tracks are required to satisfy the $\pT>1\,\GeV$ and $|\eta|<2.5$ requirements.
The following preselection requirements are used to remove combinatorial background from the sample:
 
\begin{itemize}
\item
$\chi^2(\Bcp)/\ndf < 2$, where $\chi^2(\Bcp)$ is the quality of the \Bcp cascade vertex
fit and $\ndf=8$.
 
\item
$\Lxy(\Bcp)>0.3$\,mm, where $\Lxy(\Bcp)$ is the
transverse distance between the \Bcp candidate production vertex, which corresponds to
the primary vertex of the \pp interaction (PV), and its decay vertex
projected onto the direction of the \Bcp transverse momentum.
As an event can contain several PVs, the actual PV is chosen
as the one
giving the smallest three-dimensional impact parameter of the \Bcp candidate.
To avoid biasing $\Lxy$, the PV position is recalculated after removing any tracks used in
the reconstruction of the \Bcp meson candidate.
To remove poorly reconstructed candidates, $\Lxy(\Bcp)$ is also required to not exceed 10~mm.

\item
$\Lxy(\Ds)>0$\,mm, where $\Lxy(\Ds)$ is the
transverse distance between the \Bcp decay vertex and the
\Ds decay vertex projected onto the direction of the \Ds transverse momentum.
To remove poorly reconstructed candidates, $\Lxy(\Ds)$ is also required to not exceed 10~mm.
 
\item
$|d_0^{\PV}(\Bcp)/\sigma_{d_0^{\PV}}(\Bcp)|<5$ and $|z_0^{\PV}(\Bcp)/\sigma_{z_0^{\PV}}(\Bcp)|<5$,
where $d_0^{\PV}$ and $z_0^{\PV}$ are the transverse and longitudinal
impact parameters with respect to the PV, and $\sigma_{d_0^{\PV}}(\Bcp)$ and $\sigma_{z_0^{\PV}}(\Bcp)$ are their
respective uncertainties. The uncertainties are calculated from the covariance matrix of the PV
and the covariance matrix of the $\Bcp$ pseudo-track extracted from the cascade vertex fit.
These two requirements are used to ensure that the \Bcp candidate points back to the PV.
 
\item
$\pT(\Bcp)/\sum \pT(\trk)>0.10$, where the sum is taken over all tracks originating from the selected PV,
including the tracks of the \Bcp candidate.
Due to the characteristic hard fragmentation of $b$-quarks,
this requirement removes a sizeable fraction of the combinatorial background while having almost no
effect on the signal.
\end{itemize}
 
Among various possible exclusive background contributions from
$B\to\Jpsi X$ decays, the only significant one was found to arise from
the $B_s^0\to\Jpsi\phi$ process.
Combining the tracks from a true
$B_s^0\to\Jpsi\phi$, $\Jpsi\to\mumu$, $\phi\to K^+K^-$ decay with another, random, track
may produce a fake \Bcp candidate.
To suppress background events of this type,
the \BcJpsiDsOrStar candidates with $5340<m(\Jpsi\phi)<5400$\,MeV are rejected.
 
To further suppress the combinatorial background, a multivariate classifier based on
boosted decision trees (BDTs) as implemented in the TMVA framework~\cite{TMVA} is employed.
The input variables used to train the classifier
are the \pT of the \Ds meson candidate, the $\Lxy(\Ds)$ variable, and four angular variables:
\begin{itemize}
\item
$\cos\theta^*(\pip)$, where
$\theta^*(\pip)$ is the angle between the pion momentum
in the $K^+K^-\pi^+$ rest frame and the $K^+K^-\pi^+$ combined momentum in the laboratory frame.
The signal distribution of $\cos\theta^*(\pip)$ is flat before kinematic selection because the pseudoscalar
\Ds meson decays isotropically, but it increases as $\cos\theta^*(\pip)$ approaches $+1$ for the background events.
 
\item
$|\cos^3\theta^\prime(K^+)|$, where
$\theta^\prime(K^+)$ is the angle between one of the kaons and the pion in the $K^+K^-$ rest frame.
The decay of the pseudoscalar \Ds meson to the $\phi$ (vector) plus $\pip$ (pseudoscalar)
final state results in the spin of the $\phi$ meson being aligned perpendicularly to the direction of
motion of the $\phi$ relative to the \Ds. Consequently, the distribution of $\cos\theta^\prime(K)$
follows a $\cos^2\theta^\prime(K)$ shape, implying a uniform distribution for $\cos^3\theta^\prime(K)$.
In contrast, the $\cos\theta^\prime(K)$ distribution of the combinatorial background is
approximately uniform
and its $\cos^3\theta^\prime(K)$ distribution peaks at zero.
 
\item
$\cos\theta^*(\Ds)$, where
$\theta^*(\Ds)$ is the angle between the $\Ds$ momentum
in the $B_c^+$ rest frame and the $B_c^+$ flight direction in the laboratory frame.
The distribution of $\cos\theta^*(\Ds)$ is uniform for the decays of pseudoscalar \Bcp
mesons before kinematic selection, while it tends to increase towards negative values for the background.
 
\item
$\cos\theta^\prime(\pip)$, where
$\theta^\prime(\pip)$ is the angle between the $\Jpsi$ momentum
and the pion momentum in the $K^+K^-\pip$ rest frame.
Its distribution is nearly uniform for the signal processes
but peaks towards $-1$ and $+1$ for the background.
\end{itemize}
 
The information about the helicity in the \BcJpsiDsStar decay is encoded in
the \Jpsi\Ds mass spectrum and in the distribution of
\ctpmu, where $\theta^\prime(\mup)$ is the helicity angle
defined in the rest frame of the muon pair as the angle between the
$\mu^+$ and \Ds candidate momenta. It was verified that neither $m(\Jpsi\Ds)$ nor \ctpmu
is significantly correlated with the BDT input variables.
 
The BDT classifier is trained using signal candidates from the simulated samples and background
candidates from the sidebands of the $\Jpsi\Ds$ system's mass distribution, defined as the union of the
[5680, 5900]~\MeV\ and [6400, 6800]~\MeV\ ranges. No significant difference in the BDT output is observed
when it is trained using either the \BcJpsiDs or \BcJpsiDsStar signal sample,
so the classifier used for the following selection is trained using both.
 
The ranges of the mass sidebands used to train the BDT were varied to ensure that the classifier
does not produce a bias in the mass distribution: the training is repeated using either the inner or the outer halves of
both sidebands, as well as using only the right sideband.
Alternatively, the classifier trained with the nominal sidebands was
applied to the \BcJpsiDsOrStar candidates with $\chi^2(\Bcp)/\ndf > 4$, where the signal contribution is expected to be negligible.
No significant shaping of the $m(\Jpsi\Ds)$
distribution was found in these tests.
 
The cutting threshold set on the BDT output is chosen to maximize the significance $S/\sqrt{S+B}$, where
$S$ and $B$ are the expected yields of signal and background events in the range
$5900\,\MeV < m(\Jpsi\Ds) < 6350\,\MeV$. Following the optimization procedure, a threshold
value corresponding to the efficiency of 81\% is chosen.
 
Candidates within the mass range $5680\,\MeV < m(\Jpsi\Ds) < 6800\,\MeV$ are retained for further analysis.
If more than one candidate in an event passes the selection
(this happens in about 10\% of events with at least one passing candidate), all of them are kept.

\subsection{\BcJpsiPi candidates}
\label{sec:reco:BcJpsiPi}
 
To reconstruct the \BcJpsiPi candidates, a \Jpsi candidate is combined with an additional charged-particle track
which is assigned the pion mass.
For the pion candidate, tracks identified as muons passing the \emph{low-\pT} identification working point~\cite{MUON-2018-03} are vetoed in
order to suppress a substantial background from $\Bcp\to\Jpsi\mu^+\nu_\mu X$ decays.
The three-track combination is re-fitted to a common vertex, with the \Jpsi candidate mass
constrained to the world average value~\cite{pdg2020}.
 
In what follows, the \Bcp invariant mass,
$\pT(\Bcp)$ and $\eta(\Bcp)$ are calculated using the track parameters from
the \Bcp candidate vertex fit.
The \Bcp candidates are required to have $\pT(\Bcp)>15\,\GeV$ and $|\eta(\Bcp)|<2.0$.
The re-fitted muon tracks must have $\pT>4\,\GeV$ and $|\eta|<2.3$,
while the re-fitted pion track is required to satisfy $\pt>3.5\,\GeV$ and $|\eta|<2.5$.
To suppress combinatorial background the following requirements are used:
 
\begin{itemize}
\item
$\chi^2(\Bcp)/\ndf < 1.8$ ($\ndf = 4$ in the \BcJpsiPi vertex fit),
 
\item
$\Lxy(\Bcp)>0.3$~mm,
 
\item
$|d_0^{\PV}(\Bcp)/\sigma_{d_0^{\PV}}(\Bcp)|<3$ and $|z_0^{\PV}(\Bcp)/\sigma{z_0^{\PV}}(\Bcp)|<3$,
 
\item
$\pT(\Bcp)/\sum \pT(\trk)>0.10$,
\end{itemize}
where the definitions of all these variables for the \BcJpsiPi candidates are the same as the corresponding
ones for the \BcJpsiDsOrStar selection.
 
Candidates within the mass range $5800\,\MeV < m(\Jpsi\pip) < 7100\,\MeV$ are retained for further analysis.
If more than one candidate in an event passes the selection, all of them are kept.
 
\section{Signal parameters extraction}
\label{sec:fit}
 
As described in Section~\ref{sec:detector}, various types of trigger selection chains were used to collect the events for the analysis.
Most of them are based on identification of muons from $\Jpsi\to\mumu$ decays and hence
are equally efficient for \BcJpsiDsOrStar and \BcJpsiPi decay events.
However, one suite of triggers attempts to reconstruct $\mumu\phi(K^+K^-)$ candidates; these triggers,
originally intended for the collection of $B^0_s\to\mumu\phi$ events~\cite{TRIG-2018-01}, are not efficient
for the reference decay events and may produce a bias in measurements using this channel.
Nevertheless, they contribute significantly to the \BcJpsiDsOrStar signal yields
and can be used to measure the quantities related exclusively to these decays, namely the
ratio of branching fractions \RDsStarDs and the transverse polarization fraction~\Gpp.
 
To that end, all selected $\BcJpsiDsOrStar$ candidates are separated into two (non-overlapping) datasets:
\begin{itemize}
\item \emph{Dataset 1}: candidates in the events collected by the standard dimuon triggers or by three-muon triggers without requirements on the additional ID tracks.
\item \emph{Dataset 2}: candidates collected only by the dedicated $B^0_s\to\mumu\phi$ triggers and not by other ones used in the analysis.
\end{itemize}
The \RDsPi and \RDsStarPi ratios are measured using only the \BcJpsiDsOrStar decay yields in Dataset~1,
while the full dataset, i.e. the union of Dataset~1 and Dataset~2, is employed for the \RDsStarDs and \Gpp measurements.
In the \BcJpsiPi signal yield extraction, the events collected by the $B^0_s\to\mumu\phi$ triggers
are not used at all.
 
\subsection{Extraction of \BcJpsiDsOrStar signal parameters}
\label{sec:fit:BcJpsiDs}

An extended unbinned maximum-likelihood fit to the two-dimensional distribution
of $m(\Jpsi\Ds)$ and \ctpmu is performed to extract the signal
yields as well as the transverse polarization fraction in \BcJpsiDsStar decays.
The signal and background probability density functions (PDFs) for the fit are assumed to be uncorrelated
in $m(\Jpsi\Ds)$ and \ctpmu and can therefore be factorized into
mass and angular parts.
For the signals, this assumption was validated using the
simulated samples.
A small correlation ($\sim1$\%) was observed for the background, manifesting itself
in a statistically insignificant difference between
the \ctpmu shapes of the left and right mass-sideband events;
its effect is accounted for by assigning a systematic uncertainty.
 
As stated in Section~\ref{sec:intro}, the \BcJpsiDsStar decay can be described in terms of
three helicity amplitudes: \APP, \Azz, and \AMM.
The $m(\Jpsi\Ds)$ and \ctpmu spectra should be the same for the \APP and \AMM
amplitudes, and that is confirmed with the MC simulation.
Therefore, the \BcJpsiDsStar signal is described by a model with two helicity
components corresponding to the \App and \Azz contributions.
 
Dataset~1 and Dataset~2 are fitted simultaneously.
The MC simulation indicates that both the $m(\Jpsi\Ds)$ and \ctpmu distributions of
the \BcJpsiDs signal and of the helicity components of the \BcJpsiDsStar signal are
consistent between Dataset~1 and Dataset~2. Efficiency ratios between these signal components
are consistent between Datasets~1 and~2 as well.
Therefore, the signal shapes and all signal parameters (except the yields) extracted from the fit are treated as being the same in
Dataset~1 and Dataset~2.

The $m(\Jpsi\Ds)$ and \ctpmu shapes of the two helicity components of the \BcJpsiDsStar signal are described
using templates made from the MC simulated event samples with the adaptive kernel estimation technique~\cite{Cranmer:2000du}.
The \App component's fraction in the total \BcJpsiDsStar reconstructed yield,
\fpp, is a free parameter of the fit.
 
The $m(\Jpsi\Ds)$ PDF for the \BcJpsiDs signal is parameterized
by a modified Gaussian function, $G_\textrm{mod}$~\cite{Chekanov:2005cf,BPHY-2011-02}:
\begin{equation}
G_\textrm{mod} \propto \exp\left(-0.5\times t^{\displaystyle 1+1/(1+t/2)}\right),
\label{eq:gausmod}
\end{equation}
where $t=|m(\Jpsi\Ds)-m_{\Bcp}|/\sigma_{\Bcp}$ with the mean mass $m_{\Bcp}$ and width $\sigma_{\Bcp}$ being free
parameters of the fit.
The \BcJpsiDs signal's angular PDF is also determined from the MC simulated events by using the adaptive kernel estimation technique.
 
Second-order polynomial shapes are used to describe the background \ctpmu PDFs
for Dataset~1 and Dataset~2, while the mass PDFs are
parameterized by exponential functions. The parameters of these polynomial and exponential shapes are
allowed to float independently for Dataset~1 and Dataset~2.

The free parameters of the fit also include
the ratio of the \BcJpsiDsStar yield to the \BcJpsiDs yield, $r_{\DsStar/\Ds}$,
and the \BcJpsiDs signal yields in Datasets 1 and 2,
$N_{\BcJpsiDs}^\text{DS1}$ and $N_{\BcJpsiDs}^\text{DS2}$.
The values of the signal parameters obtained from the fit to the data are shown in
Table~\ref{tab:Ds_fit_result}. It also shows the \BcJpsiDsStar signal yield in Dataset~1,
$N_{\BcJpsiDsStar}^\text{DS1}$, and the yields of both signals in the full dataset,
$N_{\BcJpsiDs}^\text{DS1\&2}$ and $N_{\BcJpsiDsStar}^\text{DS1\&2}$; their values and uncertainties
are calculated from the parameter values and covariance matrix obtained from the fit.
The fitted \Bcp mass is in good agreement with the world average value~\cite{pdg2020}
and the width agrees with the MC simulation.
The projections of the two-dimensional fit onto Dataset~1 and Dataset~2 are shown in Figure~\ref{fig:fitJpsiDs}.
 
\begin{table}[h]
\centering
\caption{
Parameters of the \BcJpsiDs and \BcJpsiDsStar signals obtained with the unbinned extended maximum-likelihood fit to the data.
Only the statistical uncertainties are included. No acceptance or efficiency corrections are applied to the signal yields.
}
\begin{tabular}{lc}
\toprule
Parameter & Value \\
\midrule
$m_{\Bcp}$ [\MeV]                  & $6274.8 \pm 1.4$  \\
$\sigma_{\Bcp}$ [\MeV]             & ~~~~$11.5 \pm 1.5$  \\
$r_{\DsStar/\Ds}$                  & ~~~~~~$1.76 \pm 0.22$  \\
$f_{\pm\pm}$                       & ~~~~~~$0.70 \pm 0.10$  \\[0.5ex]
$N_{\BcJpsiDs}^\text{DS1}$         & ~~~~$193 \pm 20$ \\[1.0ex]
$N_{\BcJpsiDs}^\text{DS2}$         & ~~~~~~$49 \pm 10$ \\
\midrule
$N_{\BcJpsiDsStar}^\text{DS1}$     & ~~~~$338 \pm 32$ \\[1.0ex]
$N_{\BcJpsiDs}^\text{DS1\&2}$      & ~~~~$241 \pm 28$ \\[1.0ex]
$N_{\BcJpsiDsStar}^\text{DS1\&2}$  & ~~~~$424 \pm 46$ \\
\bottomrule
\end{tabular}
\label{tab:Ds_fit_result}
\end{table}
 
\begin{figure}[htbp]
\centering
\subfloat[][]{
\includegraphics[width=0.49\textwidth]{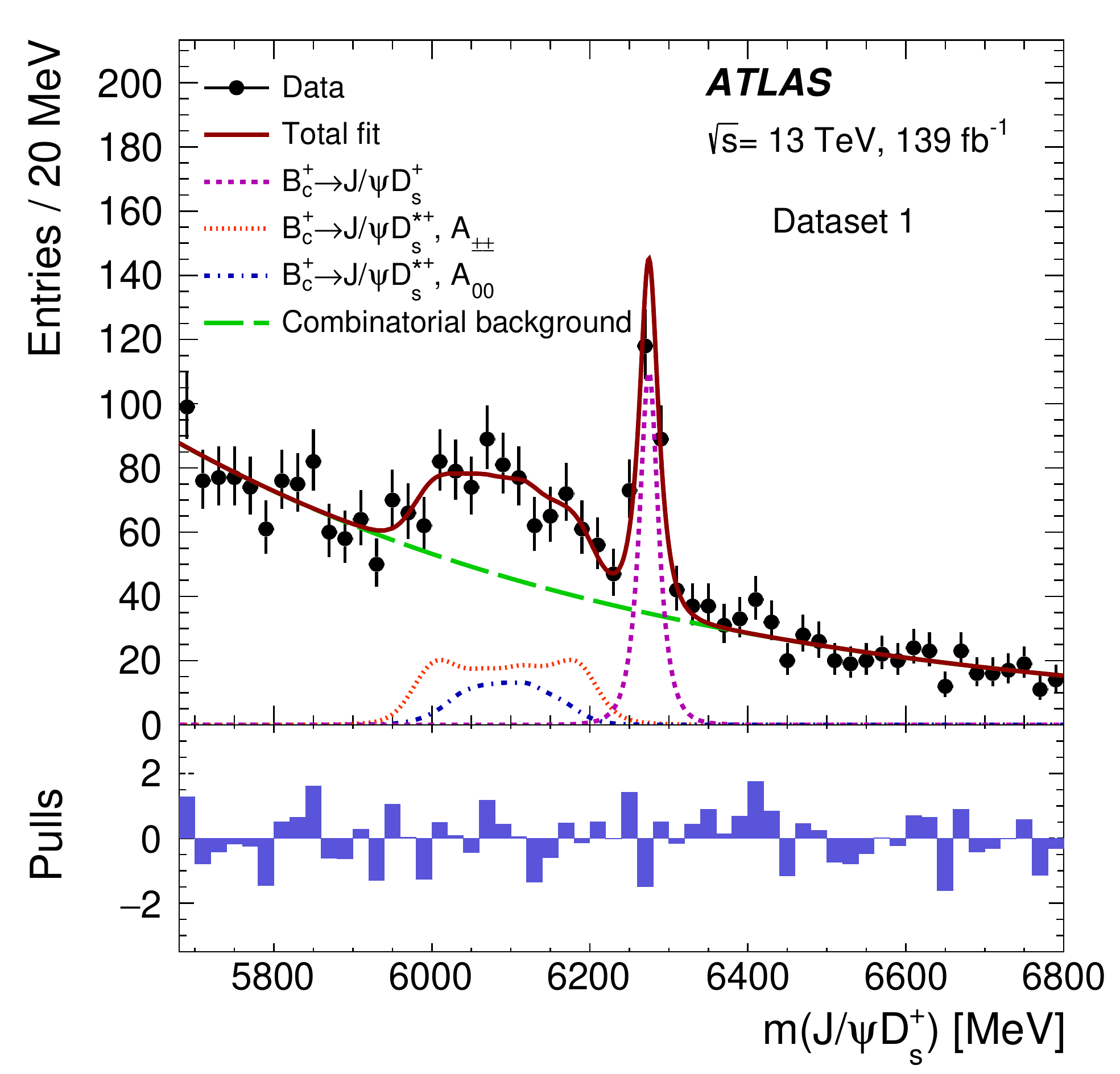}
\label{fig:fitJpsiDs:mass_DS1}
}
\subfloat[][]{
\includegraphics[width=0.49\textwidth]{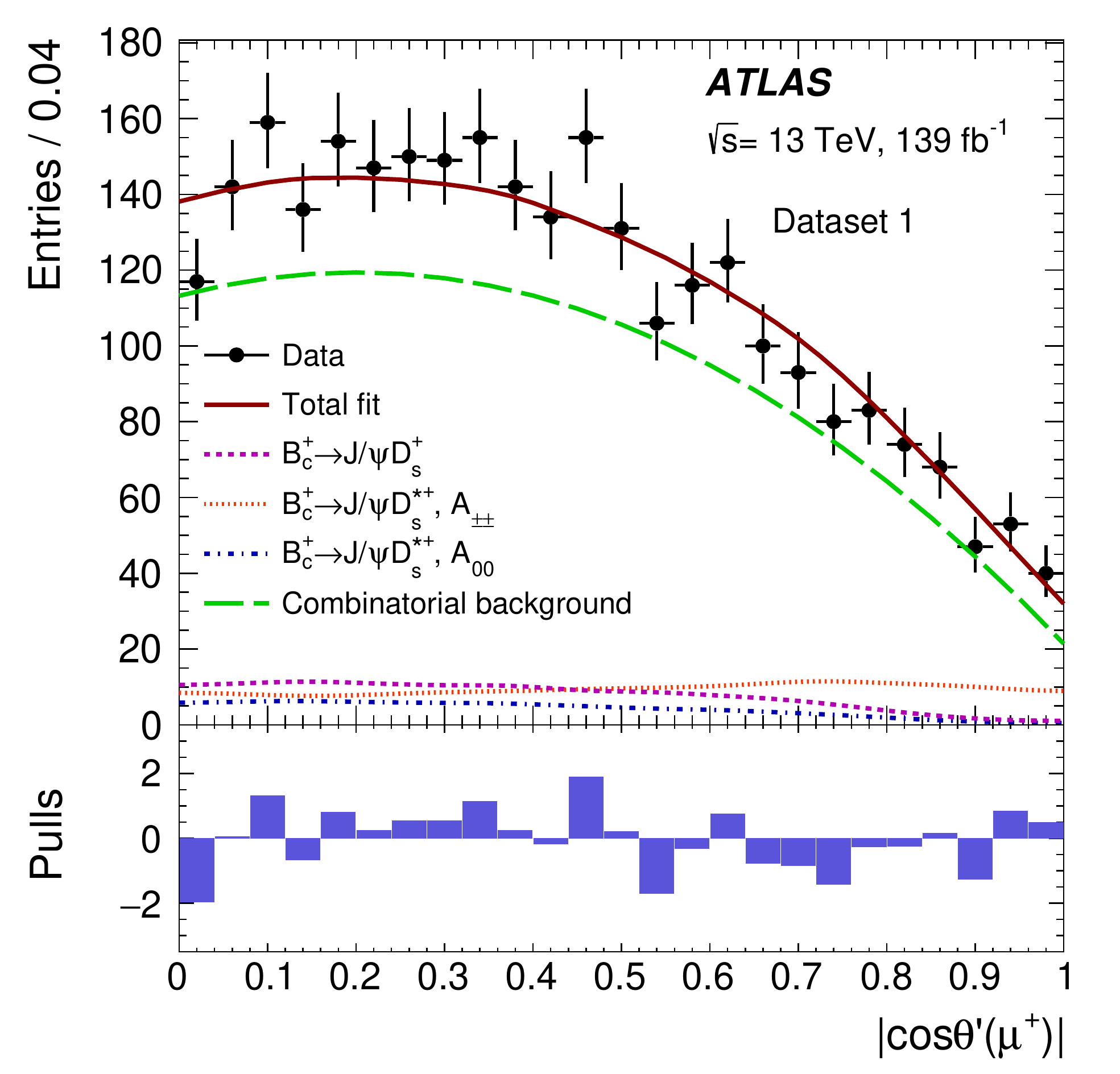}
\label{fig:fitJpsiDs:ang_DS1}
}
\\
\subfloat[][]{
\includegraphics[width=0.49\textwidth]{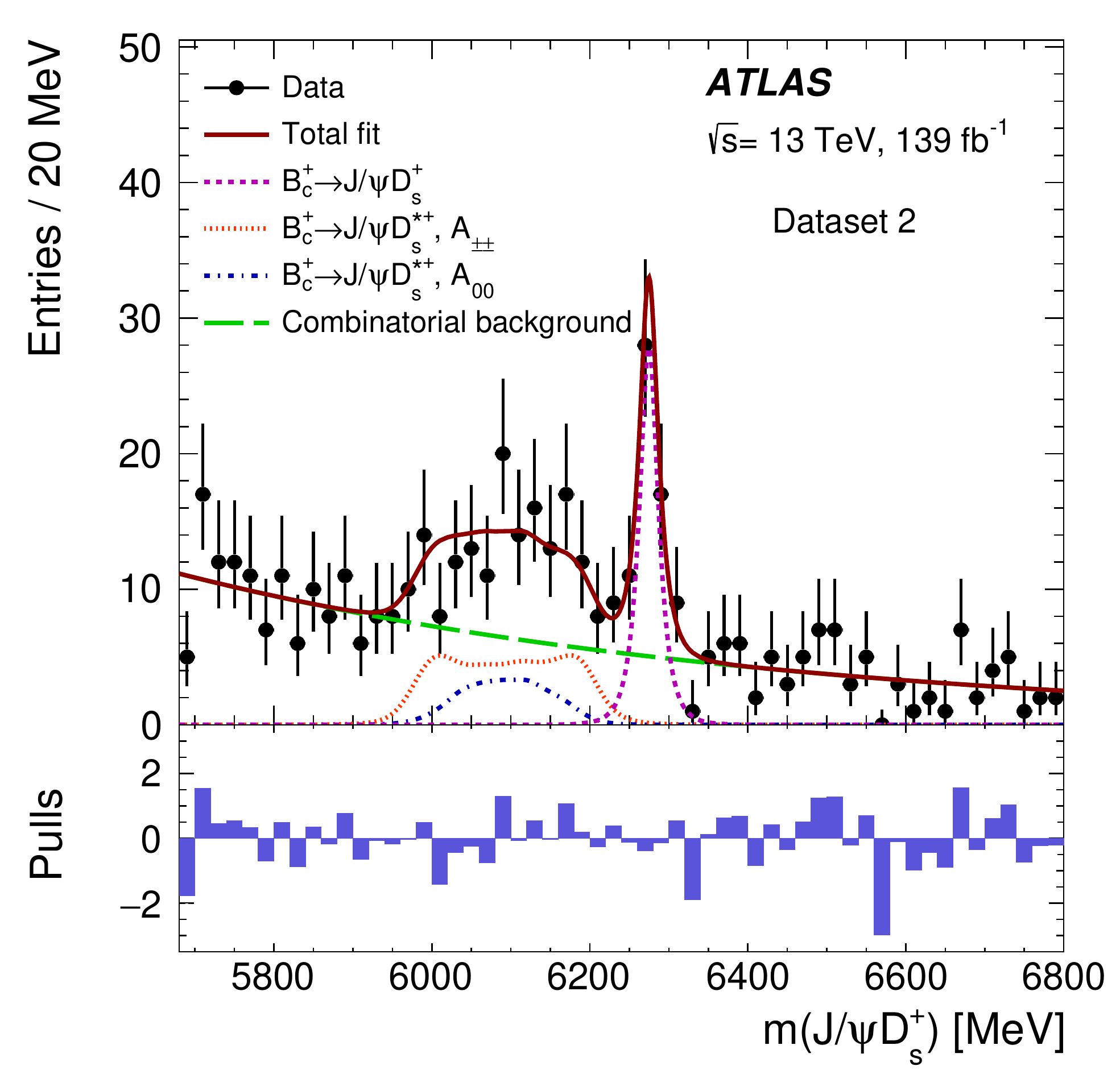}
\label{fig:fitJpsiDs:mass_DS2}
}
\subfloat[][]{
\includegraphics[width=0.49\textwidth]{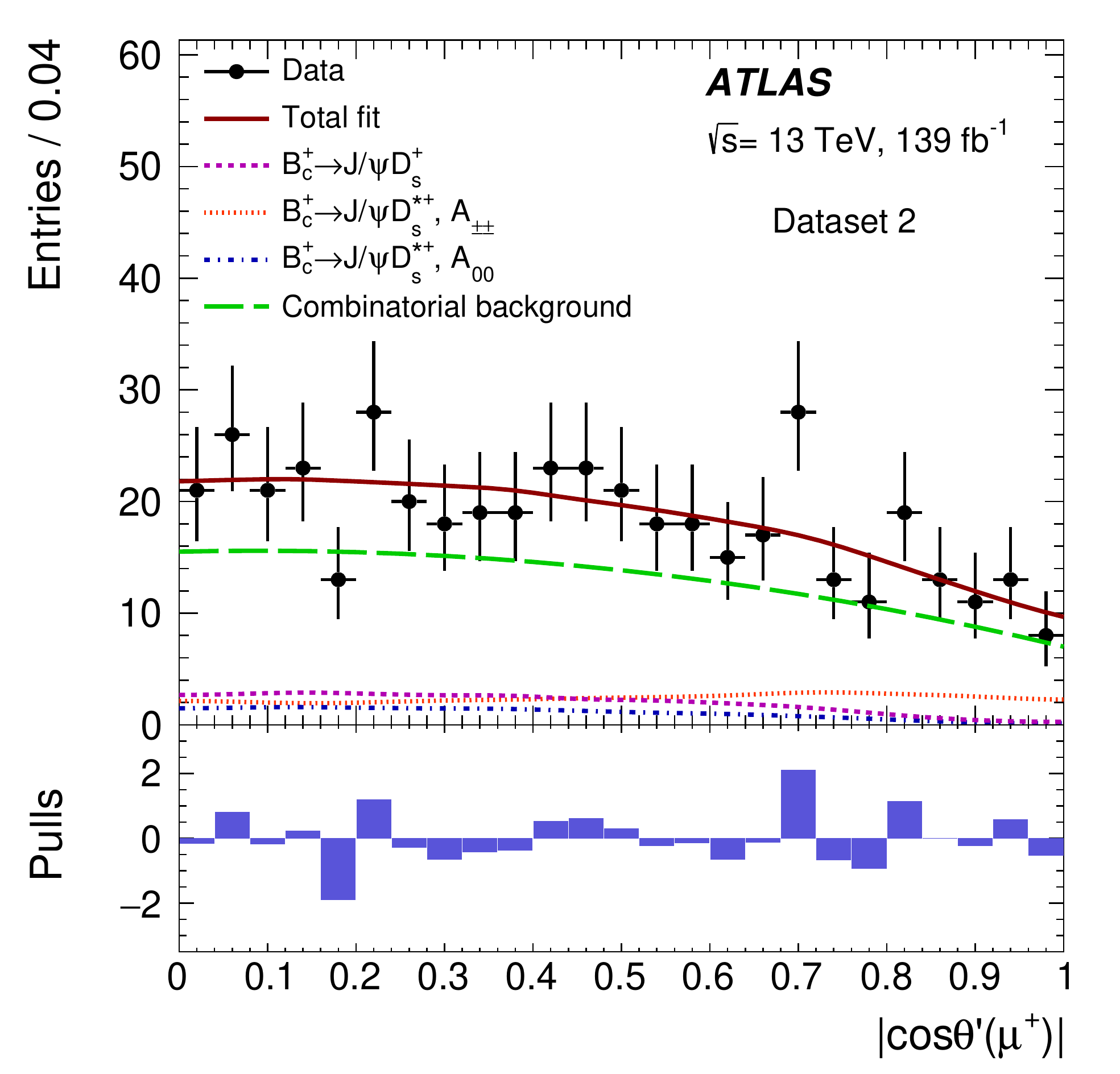}
\label{fig:fitJpsiDs:ang_DS2}
}
\caption{
Distributions of (\protect\subref{fig:fitJpsiDs:mass_DS1}, \protect\subref{fig:fitJpsiDs:mass_DS2}) the $\Jpsi\Ds$ mass and
(\protect\subref{fig:fitJpsiDs:ang_DS1}, \protect\subref{fig:fitJpsiDs:ang_DS2}) \ctpmu
for the selected \BcJpsiDsOrStar candidates in
(\protect\subref{fig:fitJpsiDs:mass_DS1}, \protect\subref{fig:fitJpsiDs:ang_DS1}) Dataset~1
and (\protect\subref{fig:fitJpsiDs:mass_DS2}, \protect\subref{fig:fitJpsiDs:ang_DS2}) Dataset~2.
Data are shown as points, and the overall result of the fit is given by the red solid curve.
The purple dashed line shows the contribution of the \BcJpsiDs signal, while the orange dotted and blue
dashed-dotted lines show the contributions of the \App and \Azz components of the \BcJpsiDsStar
signal, respectively. The background contribution is shown with the green long-dashed line.
The bottom panels show the pulls, defined as the difference between the data and the fit function
divided by the uncertainty of the data point.
\label{fig:fitJpsiDs}
}
\end{figure}

\subsection{Fit of \BcJpsiPi candidates}
\label{sec:fit:BcJpsiPi}
 
The yield of the \BcJpsiPi decay signal is extracted with
an extended unbinned maximum-likelihood fit to
the distribution of $\Jpsi\pip$ mass.
 
In the fit, the signal is modelled by a modified Gaussian function, Eq.~(\ref{eq:gausmod}), and
the combinatorial background is described
by a two-parameter exponential function, $\exp(-a_0 m\times(1 + a_1 m))$.

In addition to the combinatorial background, there is a significant contribution from partially reconstructed
\Bcp decays (PRDs), $\Bcp\to\Jpsi X$, where a charged-particle track from $X$ can be reconstructed as a pion of the \BcJpsiPi candidate.
The contribution of such decays can create structures in the left sideband of the \BcJpsiPi signal peak that could not
be described by the smooth shape of the combinatorial background PDF.
MC studies show that the dominant contribution to such decays comes from the $\Bcp\to\Jpsi\rho^+$ decay,
with $\rho^+$ decaying into $\pip\pi^0$,
while other decays, if present, exhibit a fairly smooth dependence on $m(\Jpsi\pip)$
that can be absorbed by the functional form used to model the combinatorial background.
 
The $\Bcp\to\Jpsi\rho^+$ decay is a transition of a pseudoscalar to two vector states and, similarly to \BcJpsiDsStar, can be
described in terms of three helicity amplitudes. The contributions corresponding to the $A_{\pm\pm}$ and $A_{00}$
amplitudes have a significantly different shape in $m(\Jpsi\pip)$. Therefore, to model
this background in the fit, templates for these two contributions are built out of
simulated events, using adaptive kernel estimation
technique, and added to the fit PDF, leaving their relative fraction a free parameter.
The overall normalization of this
PRD component is left free as well.
 
A small peaking background from CKM-suppressed $\Bcp\to\Jpsi K^+$ decay events,
which behaves similarly to signal, is expected in data.
The PDF for this contribution is
obtained by applying the adaptive kernel estimation technique to a simulation of the $\Bcp\to\Jpsi K^+$ channel.
The ratio of the $\Bcp\to\Jpsi K^+$ and $\Bcp\to\Jpsi\pi^+$ yields is fixed according to
the reconstruction efficiencies (nearly equal for the two modes) and the relative branching fraction~\cite{pdg2020}.
 
The mass and width of the \BcJpsiPi signal and its yield, $N_{\BcJpsiPi}$,
obtained from the fit are given in
Table~\ref{tab:pi:rslt}. The fitted \Bcp mass is in good agreement with the world average value~\cite{pdg2020}
and the width agrees with the MC simulation.
Figure~\ref{fig:mJpsipi} shows the measured $m(\Jpsi\pip)$ distribution overlaid with the result of the fit
and its signal and individual background components.
 
\begin{table}[h]
\centering
\caption{
Parameters of the \BcJpsiPi signal obtained with the unbinned extended maximum-likelihood fit.
Only the statistical uncertainties are included. No efficiency correction is applied to the signal yield.
}
\begin{tabular}{lc}
\toprule
Parameter & Value \\
\midrule
$m_{\Bcp}$ [\MeV]          & $6274.5 \pm 1.5$\\
$\sigma_{\Bcp}$ [\MeV]      & ~~~~$47.5 \pm 2.5$  \\
$N_{\BcJpsiPi}$              & ~~\,$8440 ^{+550}_{-470}$  \\
\bottomrule
\end{tabular}
\label{tab:pi:rslt}
\end{table}
 
\begin{figure}[htbp]
\centering
\includegraphics[width=0.49\textwidth]{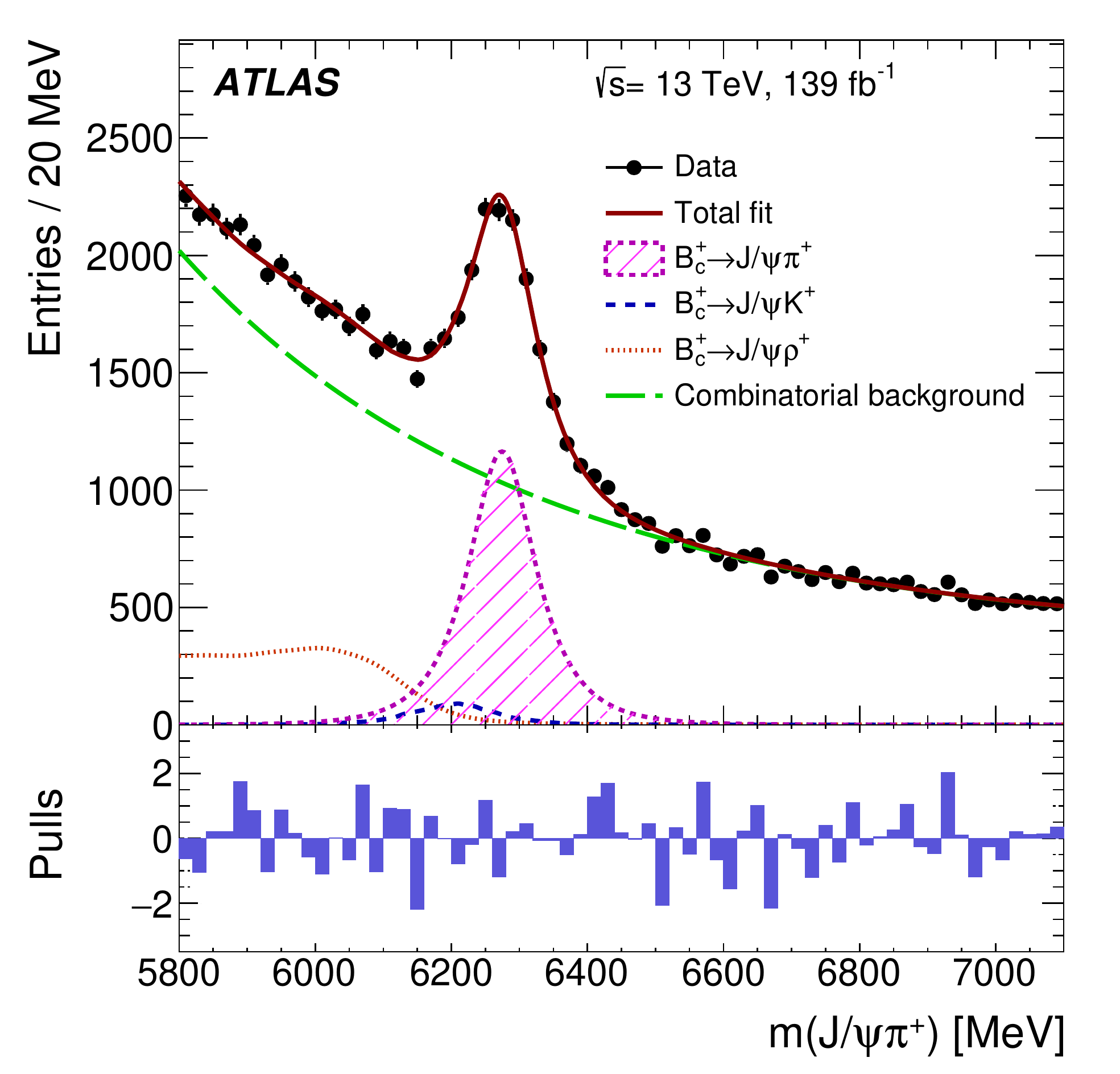}
\caption{
Distributions of the $\Jpsi\pip$ mass for the selected \BcJpsiPi candidates.
Data are shown as points, and the overall result of the fit is given by the red solid curve.
The signal contribution is shown by the purple shaded area.
The red dotted, blue dashed, and green long-dashed lines correspond to the $\Bcp\to\Jpsi\rho^+$,
$\Bcp\to\Jpsi K^+$, and combinatorial background contributions, respectively.
The bottom panel shows the pulls, defined as the difference between the data and the fit function
divided by the uncertainty of the data point.
\label{fig:mJpsipi}
}
\end{figure}

\section{Measurement of the decay parameters}
\label{sec:ratios}
The ratio of the branching fractions of \BcJpsiDsOrStar and \BcJpsiPi decays is given by
 
\begin{equation}
\label{eq:RDs}
R_{D_s^{(*)+}/\pi^+} = \frac{\mathcal B(\Bcp\to\Jpsi D_s^{(*)+})}{\mathcal B(\BcJpsiPi)}
= \frac{N_{\Bcp\to\Jpsi D_s^{(*)+}}^{\text{DS1}}}{N_{\BcJpsiPi}}\times\frac{\epsilon_{\BcJpsiPi}}{\epsilon_{\Bcp\to\Jpsi D_s^{(*)+}}^{\text{DS1}}}\times\frac{1}{\mathcal B(\Ds\to\phi(K^+K^-)\pi^+)},
\end{equation}
 
and the ratio of the branching fractions of \BcJpsiDsStar and \BcJpsiDs
is\footnote{It is assumed that $\Br(\DsStar \to \Ds X)$ is 100\%.}
 
\begin{eqnarray}
\label{eq:RDsstOverDs}
R_{\Dsst/\Ds} = \frac{\mathcal B(\BcJpsiDsStar)}{\mathcal B(\BcJpsiDs)}
= \frac{N_{\BcJpsiDsStar}^{\text{DS1\&2}}}{N_{\BcJpsiDs}^{\text{DS1\&2}}}\times\frac{\epsilon_{\BcJpsiDs}^{\text{DS1\&2}}}{\epsilon_{\BcJpsiDsStar}^{\text{DS1\&2}}}= r_{\DsStar/\Ds}\times\frac{\epsilon_{\BcJpsiDs}^{\text{DS1\&2}}}{\epsilon_{\BcJpsiDsStar}^{\text{DS1\&2}}}.
\end{eqnarray}
Here $N$ and $\epsilon$ denote the yield and the total efficiency of the corresponding mode
indicated in the subscripts.
The superscript DS1 or DS1\&2 defines whether the value corresponds to Dataset~1 or
the full dataset, respectively.
 
The total efficiency is as a product of kinematic acceptance and reconstruction efficiency.
The kinematic range of the measurement is defined in terms of the transverse momentum and pseudorapidity
of the \Bcp meson as $\pT(\Bcp) > 15$\,\GeV\ and $|\eta(\Bcp)| < 2.0$.
The kinematic acceptance is defined as the fraction of the \Bcp decay events in the kinematic
range which pass
the same requirements as those placed on the decay's muon and hadron tracks for the signal extraction.
It is determined using large generator-level samples without applying a selection on the decay products.
The reconstruction efficiency is evaluated using the MC simulation as the ratio of the number of events passing
the full analysis selection to the number of events
passing the \Bcp and decay product requirements at generator level.
The reconstruction efficiencies for Dataset~1 and for the full dataset are determined by
requiring the simulated events to pass any of the triggers qouted in the Dataset~1 definition
and any of the triggers used in the analysis, respectively.

For the value of the branching fraction $\Br(\Ds\to\phi(K^+K^-)\pi^+)$ of the $\Ds\to\phi\pip$
decay followed by $\phi\to K^+K^-$, the CLEO
measurement~\cite{Alexander:2008aa} of the partial $\Ds\to K^+K^-\pi^+$
branching fractions is used, with a kaon-pair mass within various intervals around the
nominal $\phi$ meson mass.
An interpolation between the partial branching
fractions, measured in $\pm 5$\,\MeV\ and $\pm 10$\,\MeV\ intervals by using a
relativistic Breit--Wigner distribution for the shape of the resonance, yields
the value $(1.85 \pm 0.11)$\% for the $\pm 7$\,\MeV\ interval which is used
in the analysis. This is done as in Ref.~\cite{BPHY-2014-04}.

The total efficiencies for all decay modes are shown in Table~\ref{tab:eff}.
They are different for the \App and \Azz components of the \BcJpsiDsStar decay,
hence the overall efficiency for this mode is given by
\begin{equation*}
\epsilon_{\BcJpsiDsStar} = \frac{1}{\fpp/\epsilon_{\BcJpsiDsStar, \App} + (1-\fpp)/\epsilon_{\BcJpsiDsStar, \Azz}}~,
\end{equation*}
which is valid for either Dataset~1 or the full dataset, and $\fpp$ has the value taken from the fit (shown in Table~\ref{tab:Ds_fit_result}).

\begin{table}[h]
\centering
\caption{
Summary of the total efficiencies.
Quoted uncertainties correspond to statistical uncertainties of the simulated samples.
For the \BcJpsiPi channel, the efficiency $\epsilon_{\BcJpsiPi}$ entering the equations for
\RDsOrStarPi is shown, while the efficiency for the full dataset is not defined.
}
\begin{tabular}{lcc}
\toprule
Mode        &  $\epsilon^{\text{DS1}}$ [\%] & $\epsilon^{\text{DS1\&2}}$ [\%] \\
\midrule
\BcJpsiDs   &  $0.971\pm0.012$ & $1.163\pm0.013$ \\
\BcJpsiDsStar, $\Azz$ & $0.916\pm0.012$ &  $1.088\pm0.012$ \\
\BcJpsiDsStar, $\App$    &  $0.868\pm0.010$ &  $1.049\pm0.011$ \\
\BcJpsiPi & $2.169\pm0.018$ & --              \\
\bottomrule
\end{tabular}
\label{tab:eff}
\end{table}

The fraction of transverse polarization in the \BcJpsiDsStar decay, \Gpp, is
the fraction of the $\App$ component and is calculated from the $\fpp$ value
by applying a correction to account for the difference between the total efficiencies for the
two component contributions:
\begin{equation}
\label{eq:GammaPmPm}
\Gpp = \fpp\times\frac{\epsilon_{\BcJpsiDsStar}^{\text{DS1\&2}}}{\epsilon_{\BcJpsiDsStar,\App}^{\text{DS1\&2}}}.
\end{equation}

\section{Systematic uncertainties}
\label{sec:syst}
Various sources of systematic uncertainty are considered and
outlined in this section.
The systematic uncertainties can be classified into two categories.
The first category relates to the possible differences between the data and MC simulation
which affect the signal acceptances and reconstruction efficiencies.
The second category includes the uncertainties in the signal extraction procedure,
for both the \BcJpsiDsOrStar and \BcJpsiPi channels.
Each source of systematic uncertainty is considered individually by repeating the analysis
with a certain systematic change implemented. The deviation from the nominal
result is then symmetrized and assigned as the uncertainty.
Although some sources can have rather large effects on the individual decay rate measurements,
they largely cancel out in the ratios of the branching fractions due to correlations between
their effects on the different decay modes.
 
The systematic uncertainties affecting the acceptances and reconstruction efficiencies are the following.
\begin{itemize}
\item \textit{MC modelling of \Bcp production.}
To correct for the possible difference in \Bcp kinematics between data and MC simulation,
the $\pT(\Bcp)$ and $|\eta(\Bcp)|$ spectra are extracted using the abundant \BcJpsiPi channel
with an \textit{sPlot} technique~\cite{Pivk:2004ty}.
A linear fit of data/MC ratio for these spectra is performed and weights are assigned to simulated events
according to the slope of the fitted linear function.
This slope is varied in a range preserving one standard deviation agreement with data.
The slopes are varied independently for the $\pT(\Bcp)$ and $|\eta(\Bcp)|$ distributions.
They affect both the kinematic acceptances and reconstruction efficiencies.
The effects are found to be 1.5\%--2\% for \RDsOrStarPi ratios and below 0.5\% for \RDsOrStarDs
and \Gpp.
 
\item
Limited knowledge of \textit{\Bcp and \Ds mesons lifetimes} leads to an additional systematic uncertainty due to
different decay time acceptances between the \Bcp decay modes.
The effect is estimated by varying these lifetimes within the uncertainties of
their world average values~\cite{pdg2020} and is found to be below 0.5\% for \RDsOrStarPi ratios.
 
\item The \textit{tracking efficiency uncertainty}
is dominated by the uncertainty in the detector material
description used in the MC simulation.
It is estimated from independent measurements of secondary vertices in the detector volume and results in
an effect of
about 1\% for the \RDsOrStarPi ratios and below 0.1\% for \RDsStarDs and \Gpp.
 
\item The effect of \textit{uncertainties in the modelling of multiple proton--proton interactions (pile-up)} in MC simulation is
estimated by correcting the MC pile-up profile to the profile in data. The effects on the \RDsOrStarPi ratios are
found to be about 1\%, while those on \RDsStarDs and \Gpp are statistically insignificant.
 
\item The effect of possible mis-modelling of the \textit{efficiency of $\chi^2/\ndf$ and impact parameter selection} is
evaluated by comparing their distributions in simulation and data using the \textit{sPlot}
technique, independently for \BcJpsiDsOrStar and \BcJpsiPi events. Good agreement is found,
and weights are used to vary the simulated
distribution, still preserving one standard deviation agreement with data.
The effects on the \RDsOrStarPi ratios
are found to be 3.2\% and 0.2\% for the modelling of $\chi^2/\ndf$ and the impact parameters,
respectively, with no significant effect on \RDsStarDs and \Gpp.

\item The effect of an \textit{uncertainty in the efficiency of the BDT selection} is evaluated in the same way, using
the distribution of the BDT output variable.
Very good agreement is found here as well, and variations preserving
one standard deviation agreement with data estimate an effect of 1.3\% for
\RDsOrStarPi ratios.
 
\item The possible effect of mis-modelling the \textit{trigger efficiency} is expected to be small because the same triggers
are used to select events from the different decay modes when measuring each ratio of branching fractions and \Gpp.
It is conservatively estimated using the data-driven MC corrections for dimuon triggers, applying them to all
reconstructed events and taking the deviation from the nominal result. The effect amounts to 0.6\% for \RDsOrStarPi.
For \RDsStarDs and \Gpp it is estimated at the level below 0.2\% and neglected.
The fraction of signal decay events with muons not matching those which fire the triggers is found to be about 1.5\%
and does not affect the measured quantities.
 
\item The muon reconstruction and identification efficiency uncertainty affects the individual channel
efficiencies by about 1\%--2\%. However, these mostly cancel out in the efficiency ratios and the effect
is found to be below 0.1\% for the measured quantities and is neglected.
 
\item The presence of \textit{other \Ds decay modes} with a $K^+K^-\pip$ final state affects the selection efficiency.
This is studied with a dedicated MC simulation. The dominant contribution is found to come from $\Ds\to f_0(980)(K^+K^-)\pip$ decay,
due to the large overlap of its $m(K^+K^-)$ distribution with the signal one. By considering a range of possible values for the mass and
natural width of the $f_0(980)$ meson~\cite{pdg2020} to quantify that overlap, an uncertainty of 1.6\% is assigned to the
\RDsOrStarPi ratios.

\end{itemize}
 
The following variations are used to estimate uncertainties due to the \BcJpsiDsOrStar signal fit procedure.
\begin{itemize}
\item \textit{$\Bcp\to\Jpsi D_s^{+}$ signal mass shape modelling} effects are tested with alternative
models for the \BcJpsiDs signal $m(\Jpsi\Ds)$ distribution:
a double-Gaussian function
and a double-sided Crystal Ball function~\cite{Oreglia:1980cs,Gaiser:1982yw,Skwarnicki:1986xj},
fixing the tail parameters to the values extracted from simulation.
Changes in the \BcJpsiDs yield are found to be less than 1.8\%, with smaller effects on \BcJpsiDsStar and~\fpp.
 
\item For the \textit{$\Bcp\to\Jpsi D_s^{*+}$ signal mass shape},
variations of the bandwidth scale factor in the kernel templates
($\rho$ parameter defined in Ref.~\cite{Cranmer:2000du}) in the widest range preserving both smoothness and a sufficient
level of detail in the distributions are applied.
The $m(\Jpsi\Ds)$ resolution is also varied by $\pm 10\%$.
The largest effect found is 2.7\%, for the \fpp value.
\item
The simulated \textit{shapes of the \ctpmu templates for \BcJpsiDsOrStar signals} are varied by similar changes of the
kernel bandwidth scale factor parameter.
The effect on all measured quantities is below 0.6\%.
 
\item For the \textit{background mass shape modelling}, the two-parameter exponential function
and a second-order polynomial function are used as alternatives to the simple exponential shape;
the fitted mass range is also reduced in turn by 40\,\MeV\
(two bins of the mass distributions in Figure~\ref{fig:fitJpsiDs})
from the low-mass and high-mass ends.
These variations represent the largest uncertainty in the \BcJpsiDsOrStar signal extraction and amount to 6.0\%, 9.0\%, and
3.2\% for \RDsPi, \RDsStarPi, and \RDsStarDs, respectively.
 
\item The \textit{background \ctpmu shape} is alternatively parameterized with a third-order polynomial function. A linear combination
of the kernel templates made from the left and right data sidebands is also used to describe the background in the mass range of
the signals. The latter variation also covers a small deviation from the assumption of factorization of the mass and angular
parts of the background PDF.
These variations yield a maximum change of about 2\% for \RDsStarDs and \Gpp, and about 1\% for \RDsPi and \RDsStarPi.
 
\item
Effect of the \textit{$B^0_s\to\mumu\phi$ triggers}: in the nominal fit, the angular and mass shapes of the signals are supposed to be the same in
Datasets~1 and~2, and the same value of \fpp is used in both.
To estimate the possible effects of deviations from these assumptions, alternative
fit models with either of the following changes are tested:
\begin{itemize}
\item different signal mass and \ctpmu templates are used for Datasets~1 and~2
(produced from events selected by the corresponding triggers);
\item the \fpp fraction is set to be different between Datasets~1 and~2 according to slightly different efficiencies in the simulation.
\end{itemize}
These variations yield an uncertainty of about 4\% for \Gpp and less than 1.5\% for the $R$ ratios.
 
\item The \textit{branching fractions of the \DsStar decays} are varied in simulation within their uncertainties~\cite{pdg2020} since they affect the
\BcJpsiDsStar template shapes. The effect is found to be 0.7\% for \Gpp and below 0.1\% for the other quantities.

\end{itemize}
 
The following variations are used to estimate uncertainties due to the \BcJpsiPi signal fit procedure.
\begin{itemize}
\item For the \emph{signal modelling},
a double-Gaussian function
and a double-sided Crystal Ball function are used as alternatives to the modified Gaussian function,Eq.~(\ref{eq:gausmod}),
fixing the tail parameters to the values extracted from simulation.
Changes in the signal yields reach 4.2\%.
\item Alternative models are used for \emph{combinatorial background modelling}:
a three-parameter exponential function,
and second- and third-order polynomial functions are used.
These models, providing more freedom at lower $m(\Jpsi\pip)$
values, are able to absorb a part of PRD contribution that may not have been accounted for by using the
$\Bcp\to\Jpsi\rho^+$ templates.
Alternatively, an additional hyperbolic tangent function is added to the fit PDF
to cover possible mis-modelling of the PRD component.
The shape of the $\Bcp\to\Jpsi\rho^+$ templates is also varied
by changing the kernel bandwidth scale factor and $m(\Jpsi\pip)$ resolution.
These variations produced changes of up to 5.8\% in the signal yield.
\item
For the \emph{CKM-suppressed peaking background} from $\Bcp\to\Jpsi K^+$, its fraction is varied according to
the uncertainty in the measured value of $\Br(\Bcp\to\Jpsi K^+)/\Br(\BcJpsiPi)$~\cite{pdg2020}.
The change in the signal yield is 1\%.
\end{itemize}
 
The statistical uncertainties of the total efficiency values due to the limited number of MC events are also treated as
a separate source of systematic uncertainty.
 
Contributions from the different sources of systematic uncertainty described above are summed in quadrature.
Finally, since the branching fraction of $\Ds\to\phi(K^+K^-)\pip$ is featured in Eq.~(\ref{eq:RDs}),
its uncertainty~\cite{Alexander:2008aa} is propagated
to the final values of the relative branching fractions \RDsOrStarPi.
This uncertainty is quoted separately.
 
The systematic uncertainties of the measured quantities are summarized in Table~\ref{tab:syst}.
 
\begin{table}[h]
\centering
\caption{
Relative systematic uncertainties of the measured quantities.
}
\begin{tabular}{l|cccc}
\toprule
Source & \multicolumn{4}{c}{Uncertainty [\%]} \\
& $R_{\Ds/\pi^+}$ & $R_{\Dsst/\pi^+}$ & $R_{\Dsst/\Ds}$ & $\Gamma_{\pm\pm}/\Gamma$ \\
\midrule
Simulated $\pt(\Bcp)$ spectrum           & 1.5    & 1.9     & 0.4    & 0.1     \\
Simulated $|\eta(\Bcp)|$ spectrum        & 0.7    & 0.7     & 0.1    & 0.2     \\
$\Bcp$ lifetime                          & 0.1    & $<0.1$~~~~  & --     & --      \\
$\Ds$ lifetime                           & 0.4    & 0.4     & --     & --      \\
Tracking efficiency                      & 1.0    & 1.0     & $<0.1$~~~~ & $<0.1$~~~~  \\
Pile-up effects                          & 1.0    & 1.0     & --     & --      \\
$\chi^2/\ndf$ selection efficiency       & 3.2    & 3.2     & --     & --      \\
Impact parameter selection efficiency    & 0.2    & 0.2     & --     & --      \\
BDT selection efficiency                 & 1.3    & 1.3     & --     & --      \\
Trigger efficiency                       & 0.6    & 0.6     & --     & --      \\
Other \Ds decay modes                    & 1.6    & 1.6     & --     & --      \\
\midrule
$\BcJpsiDsOrStar$ signal fit:            &        &         &        &         \\
\quad $\Ds$ signal mass modelling        & 1.8    & 0.5     & 1.3    & 0.8     \\
\quad $\DsStar$ signal mass modelling    & 0.6    & 1.2     & 1.7    & 2.7     \\
\quad Signal angular modelling           & 0.4    & $<0.1$~~~~  & 0.4    & 0.6     \\
\quad Background mass modelling          & 6.0    & 9.0     & 3.2    & 1.0     \\
\quad Background angular modelling       & 0.9    & 1.3     & 2.1    & 2.4     \\
\quad $B^0_s\to\mumu\phi$ triggers       & 0.8    & 0.5     & 1.3    & 4.0     \\
\quad \DsStar branching fractions        & $<0.1$~~~~ & $<0.1$~~~~  & $<0.1$~~~~ & 0.7     \\
\BcJpsiPi signal fit:                    &        &         &        &         \\
\quad Signal modelling                   & 4.2    & 4.2     & --     & --      \\
\quad PRD/comb.\ background modelling     & 5.8    & 5.8     & --     & --      \\
\quad CKM-suppr.\ background modelling    & 1.0    & 1.0     & --     & --      \\
\midrule
MC statistics                            & 1.5    & 1.5     & 1.7    & 1.5     \\
\midrule
Total                                    & 10.7~~   & 12.6~~    & 4.9    & 5.8     \\
\midrule
$\Br(\Ds\to\phi(K^+K^-)\pip)$            & 5.9    & 5.9     & --     & --      \\
\bottomrule
\end{tabular}
\label{tab:syst}
\end{table}

\section{Results and discussion}
\label{sec:result}
 
The ratios of the branching fractions for \BcJpsiDsOrStar and \BcJpsiPi
calculated according to Eq.~(\ref{eq:RDs}) are found to be
\begin{equation*}
\RDsPi = 2.76\pm 0.33\pm 0.29\pm 0.16
\end{equation*}
and
\begin{equation*}
\RDsStarPi = 5.33\pm 0.61\pm 0.67\pm 0.32,
\end{equation*}
where the first uncertainty is statistical, the second is systematic, and the third corresponds to the uncertainty in
the branching fraction of the $\Ds\to\phi(K^+K^-)\pip$ decay.
The ratio of the branching fractions for \BcJpsiDsStar and \BcJpsiDs
calculated according to Eq.~(\ref{eq:RDsstOverDs}) is found to be
\begin{equation*}
\RDsStarDs = 1.93\pm0.24 \pm 0.09.
\end{equation*}
The fraction of transverse polarization in \BcJpsiDsStar
decay calculated according to Eq.~(\ref{eq:GammaPmPm}) is found to be
\begin{equation*}
\Gpp = 0.70\pm0.10\pm 0.04.
\end{equation*}
In the last two quantities, the first uncertainty is statistical and the second is systematic.
 
These ATLAS results from \RunTwo are compared with the results of LHCb~\cite{LHCb:2013kwl} and
ATLAS~\RunOne~\cite{BPHY-2014-04} measurements and with the predictions from
various theoretical calculations in Table~\ref{tab:PredicRelBr} and Figure~\ref{fig:ValComp}.
The measurements of \RDsPi and \RDsStarDs agree with the ATLAS~\RunOne and LHCb results.
The obtained values of \RDsStarPi and \Gpp are, respectively, smaller and larger than in the ATLAS~\RunOne measurement,
although in each case the difference does not
exceed 1.5 standard deviations of the \RunOne uncertainty.
Since the precision of the new results significantly exceeds that
achieved in the ATLAS \RunOne analysis~\cite{BPHY-2014-04},
the latter is superseded by this work.
 
\begin{table}[ht]
\begin{center}
\caption{Comparison of the measured quantities with the results of earlier measurements and theory predictions.
Values of the corresponding ratios of branching fractions calculated using
Eqs.~(\ref{eq:Factorization:RDsPi})--(\ref{eq:Factorization:RDsStarDs}) and
Eq.~(\ref{eq:Factorization:RDsStarDs_sup})
and of transverse polarization fractions for $B^+$, $B^0$, and $B^0_s$ decays are also shown.
No phase-space corrections are applied to the ratios
and the quoted uncertainties are the ones propagated from the world average uncertainties
of the individual decay branching fractions.
For all experimental measurements, quadrature sums of all uncertainties are quoted.
Uncertainties in the predictions are quoted only if they are explicitly given in the corresponding references.
\label{tab:PredicRelBr}}
\begin{tabular}{c c c c r}
\toprule
\RDsPi  & \RDsStarPi  & \RDsStarDs  & \Gpp  & Reference \\
\midrule
$2.76\pm 0.47$ & $5.33\pm 0.96$ & $1.93\pm0.26$ & $0.70\pm0.11$ & ATLAS \RunTwo \\
\midrule
$2.90\pm 0.62$ & -- & $2.37\pm 0.57$ & $0.52 \pm 0.20$ & LHCb \RunOne~\cite{LHCb:2013kwl} \\
$3.8 \pm 1.2$ & $10.4 \pm 3.5$~~ & $2.8^{+1.2}_{-0.9}$\, & $0.38 \pm 0.24$ & ATLAS \RunOne~\cite{BPHY-2014-04} \\
\midrule
$2.6$ & $4.5$ & $1.7$ & -- & QCD potential model \cite{Colangelo:1999zn} \\
$1.3$ & $5.2$ & $3.9$ & -- & QCD sum rules \cite{Kiselev:2002vz} \\
$1.29\pm 0.26$ & $5.09\pm 1.02$ & $3.96\pm 0.80$ & $0.46\pm 0.09$ & CCQM \cite{Dubnicka:2017job} \\
 
$2.2$ & -- & -- & -- & BSW \cite{Dhir:2008hh} \\
$2.06\pm 0.86$ & -- & $3.01\pm 1.23$ & -- & LFQM \cite{Ke:2013yka} \\
$3.45^{+0.49}_{-0.17}$~ & -- & $2.54^{+0.07}_{-0.21}$~\, & $0.48\pm 0.04$ & pQCD \cite{Rui:2014tpa} \\
$3.7832$ & -- & -- & $0.410$ & RIQM \cite{Kar:2013fna, Nayak:2022qaq} \\
$3.257\pm 0.293$ & -- & -- & -- & FNCM \cite{Mohammadi:2018jcp}\\
\midrule
$1.67 \pm 0.36$ & $3.49 \pm 0.52$ & $2.09 \pm 0.52$ & -- & $B^+\to\bar D^{*0}D_s^{(*)+}/\bar D^{*0}\pip$~\cite{pdg2020} \\
$2.92 \pm 0.42$ & $6.46 \pm 0.60$ & $2.21 \pm 0.35$ & $0.48 \pm 0.05$ & $B^0\to D^{*-}D_s^{(*)+}/D^{*-}\pip$~\cite{pdg2020} \\
-- & $7.2  \pm 2.1 $ & -- & $0.94 \pm 0.18$ & $B^0_s\to D_s^{*-}D_s^{+}/D_s^{*-}\pip$~\cite{pdg2020} \\
\midrule
-- & -- & $1.402 \pm 0.083$ & $0.396 \pm 0.023$ & $B^+\to\Jpsi K^{(*)+}$~\cite{pdg2020} \\
-- & -- & $1.425 \pm 0.065$ & $0.429 \pm 0.007$ & $B^0\to\Jpsi K^{(*)0}$~\cite{pdg2020} \\
-- & -- & --    & $0.4774 \pm 0.0034$ & $B^0_s\to\Jpsi \phi$~\cite{pdg2020} \\
\bottomrule
\end{tabular}
\end{center}
\end{table}
 
\begin{figure}[htbp]
\centering
\includegraphics[width=0.9\textwidth]{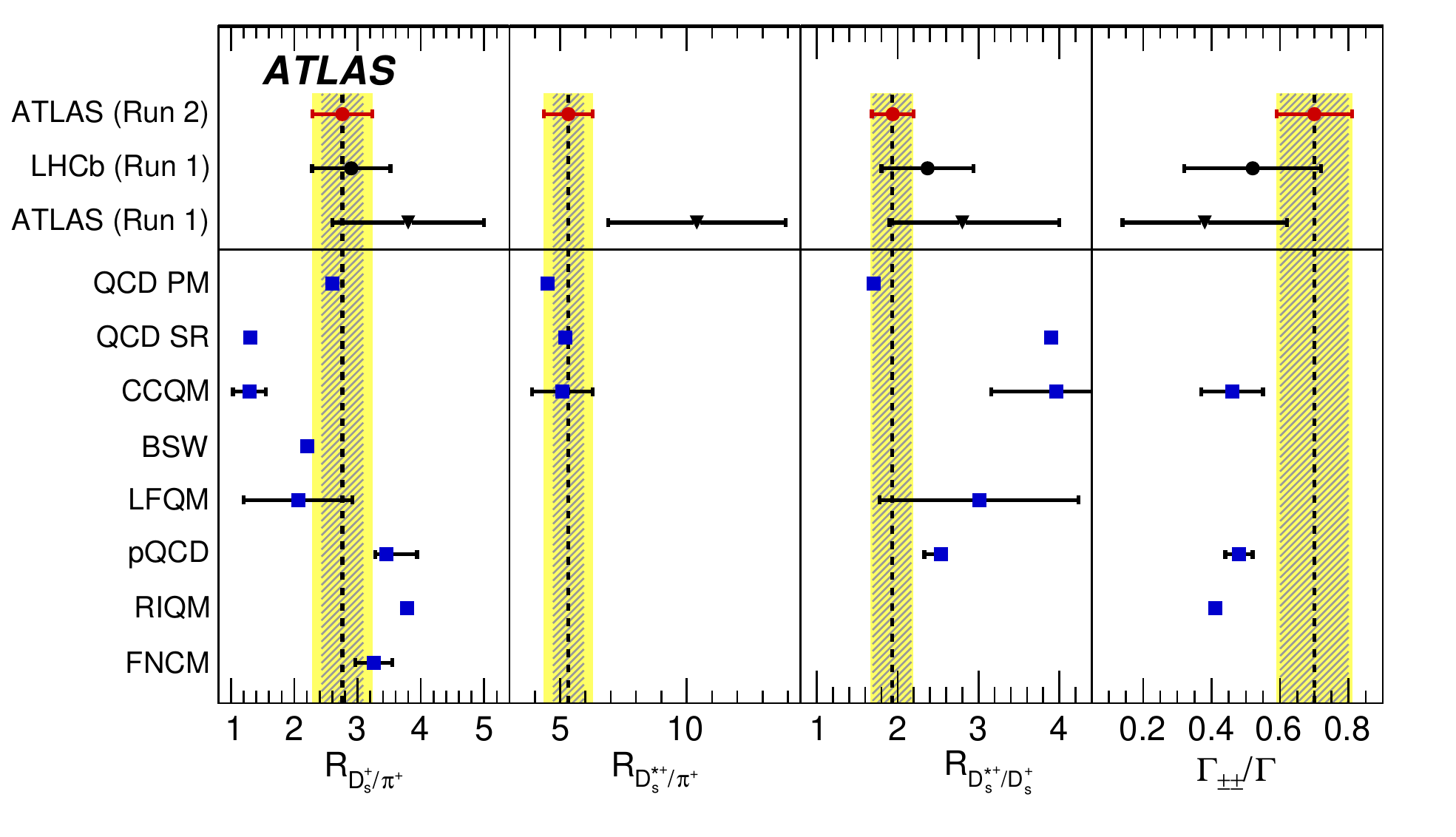}
\caption{Comparison of the results of this measurement with those of ATLAS \RunOne~\cite{BPHY-2014-04},
LHCb~\cite{LHCb:2013kwl} and theoretical predictions based
on a QCD relativistic potential model (QCD PM)~\cite{Colangelo:1999zn},
QCD sum rules (QCD SR)~\cite{Kiselev:2002vz},
covariant confined quark model (CCQM)~\cite{Dubnicka:2017job},
Bauer--Stech--Wirbel relativistic quark model (BSW)~\cite{Dhir:2008hh},
light-front quark model (LFQM)~\cite{Ke:2013yka},
perturbative QCD (pQCD)~\cite{Rui:2014tpa},
relativistic independent quark model (RIQM)~\cite{Kar:2013fna, Nayak:2022qaq},
and calculations in the QCD factorization approach (FNCM)~\cite{Mohammadi:2018jcp}.
Hatched areas show the statistical uncertainties of this measurement and yellow bands correspond to the total uncertainties.
Quadrature sums of all experimental uncertainties are quoted for the ATLAS \RunOne and LHCb results.
The uncertainties of the theoretical predictions are shown only if they are explicitly quoted in the corresponding papers.
\label{fig:ValComp}
}
\end{figure}
 
The ratios of branching fractions are well described by the predictions of a QCD relativistic potential model~\cite{Colangelo:1999zn}.
The predictions of a QCD sum rules calculation~\cite{Kiselev:2002vz}, the
covariant confined quark model~\cite{Dubnicka:2017job}\footnote{
The predictions of relativistic quark model~\cite{Ivanov:2006ni} are made within a very similar approach, so only the updated result
from Ref.~\cite{Dubnicka:2017job} is discussed.
}
and the light-front quark model~\cite{Ke:2013yka}
all underestimate the \RDsPi ratio, although the prediction of the latter model still agrees with data within a large theoretical uncertainty.
However, the \RDsStarPi ratio is described well by the same predictions, which correspondingly overestimate the \RDsStarDs ratio.
Nonetheless, for the light-front quark model the predicted value of \RDsStarDs is still in agreement within the theoretical uncertainty.
 
The \RDsPi value predicted in the Bauer--Stech--Wirbel framework~\cite{Dhir:2008hh}
is slightly below the data, while the results of calculations within
perturbative QCD~\cite{Rui:2014tpa} and the QCD factorization approach~\cite{Mohammadi:2018jcp} are near the upper bound of the
experimental uncertainty range;
the recent prediction made within the framework of the relativistic independent quark model~\cite{Nayak:2022qaq} gives an
even higher value.
The perturbative QCD prediction~\cite{Rui:2014tpa} for the \RDsStarDs ratio is also slightly above the data.

The measured value of \Gpp clearly agrees with a naive spin-counting expectation of $2/3$.
It is larger than the values calculated in the covariant confined quark model~\cite{Dubnicka:2017job}, perturbative QCD~\cite{Rui:2014tpa}, and
the relativistic independent quark model~\cite{Kar:2013fna}, which give values below 0.5.
The discrepancies, however, do not exceed 2--3 standard deviations of the experimental uncertainty.
 
The measured ratios of branching fractions and the transverse polarization fraction can be compared with
the corresponding quantities for the other $B$ mesons. Assuming that the colour-favoured spectator
diagram (shown in Figure~\ref{fig:Diagrams:Diag_spectator} for \BcJpsiDsOrStar) dominates the decay amplitudes, the following approximate
relations can be established:
\begin{equation}
\label{eq:Factorization:RDsPi}
\RDsPi \approx
\frac{\Gamma(B\to \bar{D}^*\Ds)}{\Gamma(B\to \bar{D}^*\pi^+)},
\end{equation}
\begin{equation}
\label{eq:Factorization:RDsStarPi}
\RDsStarPi \approx
\frac{\Gamma(B\to
\bar{D}^*\DsStar)}{\Gamma(B\to \bar{D}^*\pi^+)},
\end{equation}
\begin{equation}
\label{eq:Factorization:RDsStarDs}
\RDsStarDs \approx
\frac{\Gamma(B\to
\bar{D}^*\DsStar)}{\Gamma(B\to \bar{D}^*\Ds)},
\end{equation}
where $B$ and $\bar{D}^*$ stand for $B^+$ and $\bar{D}^{*0}$, $B^0$ and $D^{*-}$, or $B^0_s$ and $D_s^{*-}$.
Similarly, the \Gpp value for \BcJpsiDsStar decay can be compared with the same
quantity in the corresponding $B^+$ or $B^0_s$ decays.
 
One can also compare the measured value of \RDsStarDs with the corresponding ratio for the $B^0$ and $B^+$ decays
occurring predominantly via the colour-suppressed spectator diagram:
\begin{equation}
\label{eq:Factorization:RDsStarDs_sup}
\RDsStarDs \sim
\frac{\Gamma(B\to \Jpsi K^*)}{\Gamma(B\to \Jpsi K)},
\end{equation}
where $B$ stands for $B^+$ or $B^0$ and $K^{(*)}$ is $K^{(*)+}$ or $K^{(*)0}$, respectively. The \Gpp value, in turn,
can be compared with the transverse polarization fraction in the same $B\to \Jpsi K^*$ decays as well as
in $B^0_s\to\Jpsi\phi$ decays.
 
The estimates obtained using Eqs.~(\ref{eq:Factorization:RDsPi})--(\ref{eq:Factorization:RDsStarDs}) and
Eq.~(\ref{eq:Factorization:RDsStarDs_sup}) with the world average branching fraction values for the $B^0$, $B^+$,
and $B^0_s$ decays~\cite{pdg2020} are also shown in Table~\ref{tab:PredicRelBr},
along with the \Gpp values for the corresponding decays of those mesons~\cite{pdg2020}.
Corrections to the above approximations, required due to phase-space differences between the
studied \Bc decays and the corresponding $B^+$, $B^0$, and $B^0_s$ decays, are 1\%--2\% for
Eqs.~(\ref{eq:Factorization:RDsPi})--(\ref{eq:Factorization:RDsStarDs}) and increase the values from
Eq.~(\ref{eq:Factorization:RDsStarDs_sup}) by about 7\%.
These corrections are not applied in the numbers quoted in Table~\ref{tab:PredicRelBr}.
 
The \RDsStarDs value agrees with the corresponding ratio calculated with Eq.~(\ref{eq:Factorization:RDsStarDs})
for both the $B^0$ and $B^+$ decays and is larger than that obtained from Eq.~(\ref{eq:Factorization:RDsStarDs_sup}) for
decays of the same mesons. It supports the assumption that the colour-favoured spectator diagram dominates
the \BcJpsiDsOrStar decay amplitudes. The \RDsPi and \RDsStarPi values agree with the calculations made
using Eqs.~(\ref{eq:Factorization:RDsPi})--(\ref{eq:Factorization:RDsStarPi}) for decays of $B^0$ and $B^0_s$, and are
larger than those for decays of $B^+$. The measured value of \Gpp lies between the transverse polarization
fraction values in the $B^0\to D^{*-} D_s^{*}$ and $B^0_s\to D_s^{*-} D_s^{*}$ decays and is larger than one in
the considered $B$ decays occurring via the colour-suppressed spectator diagram.

\FloatBarrier

\section{Summary}
\label{sec:summary}
A study of \BcJpsiDs and \BcJpsiDsStar decays has been performed
by the ATLAS experiment at the LHC using \pp collision
data corresponding to an integrated luminosity of $139$\,\ifb\
at 13\,\TeV\ centre-of-mass energy.
The ratios of the branching fractions of these decays to the branching fraction of
\BcJpsiPi decay are measured to be $\RDsPi = 2.76\pm 0.33\pm 0.29\pm 0.16$ and
$\RDsStarPi = 5.33\pm 0.61\pm 0.67\pm 0.32$, where the first uncertainty is statistical,
the second is systematic, and the third is due to the uncertainty in
the branching fraction of $\Ds\to\phi(K^+K^-)\pip$ decay.
The ratio of the two decay branching
fractions is also measured, yielding $\RDsStarDs = 1.93\pm0.24 \pm 0.09$.
The transverse polarization fraction in the \BcJpsiDsStar decay
is found to be $\Gpp = 0.70\pm0.10\pm 0.04$.
The results of this new measurement supersede those of the previous ATLAS measurement using \RunOne data.
 
All results are consistent with the earlier measurements by ATLAS and LHCb.
The precision of the measurement exceeds that of
all previous studies of these decays.
 
Various predictions for the measured quantities are compared with data.
A QCD relativistic potential model calculation agrees well with all three ratios of branching fractions.
The \RDsStarPi ratio also agrees with other calculations made using QCD sum rules
and the covariant confined quark model. The \RDsPi and \RDsStarDs ratios are, however,
poorly described by the same predictions.
The measured transverse polarization fraction in the \BcJpsiDsStar decay agrees well with the value of $2/3$ expected
from naive spin-counting considerations, while the available calculations of this quantity in the
covariant confined quark model, perturbative QCD, and the
relativistic independent quark model
give values below the data, although agreeing within 2--3 standard deviations of experimental uncertainty.
 
Comparisons of the measured \RDsStarDs ratio with the similar ratios for the $B^0$ and $B^+$ meson decays
support the assumption that the colour-favoured spectator diagram dominates the \BcJpsiDsOrStar decay
amplitudes.

\section*{Acknowledgements}
 

We thank CERN for the very successful operation of the LHC, as well as the
support staff from our institutions without whom ATLAS could not be
operated efficiently.
 
We acknowledge the support of
ANPCyT, Argentina;
YerPhI, Armenia;
ARC, Australia;
BMWFW and FWF, Austria;
ANAS, Azerbaijan;
SSTC, Belarus;
CNPq and FAPESP, Brazil;
NSERC, NRC and CFI, Canada;
CERN;
ANID, Chile;
CAS, MOST and NSFC, China;
Minciencias, Colombia;
MEYS CR, Czech Republic;
DNRF and DNSRC, Denmark;
IN2P3-CNRS and CEA-DRF/IRFU, France;
SRNSFG, Georgia;
BMBF, HGF and MPG, Germany;
GSRI, Greece;
RGC and Hong Kong SAR, China;
ISF and Benoziyo Center, Israel;
INFN, Italy;
MEXT and JSPS, Japan;
CNRST, Morocco;
NWO, Netherlands;
RCN, Norway;
MEiN, Poland;
FCT, Portugal;
MNE/IFA, Romania;
JINR;
MES of Russia and NRC KI, Russian Federation;
MESTD, Serbia;
MSSR, Slovakia;
ARRS and MIZ\v{S}, Slovenia;
DSI/NRF, South Africa;
MICINN, Spain;
SRC and Wallenberg Foundation, Sweden;
SERI, SNSF and Cantons of Bern and Geneva, Switzerland;
MOST, Taiwan;
TAEK, Turkey;
STFC, United Kingdom;
DOE and NSF, United States of America.
In addition, individual groups and members have received support from
BCKDF, CANARIE, Compute Canada and CRC, Canada;
COST, ERC, ERDF, Horizon 2020 and Marie Sk{\l}odowska-Curie Actions, European Union;
Investissements d'Avenir Labex, Investissements d'Avenir Idex and ANR, France;
DFG and AvH Foundation, Germany;
Herakleitos, Thales and Aristeia programmes co-financed by EU-ESF and the Greek NSRF, Greece;
BSF-NSF and GIF, Israel;
Norwegian Financial Mechanism 2014-2021, Norway;
NCN and NAWA, Poland;
La Caixa Banking Foundation, CERCA Programme Generalitat de Catalunya and PROMETEO and GenT Programmes Generalitat Valenciana, Spain;
G\"{o}ran Gustafssons Stiftelse, Sweden;
The Royal Society and Leverhulme Trust, United Kingdom.
 
The crucial computing support from all WLCG partners is acknowledged gratefully, in particular from CERN, the ATLAS Tier-1 facilities at TRIUMF (Canada), NDGF (Denmark, Norway, Sweden), CC-IN2P3 (France), KIT/GridKA (Germany), INFN-CNAF (Italy), NL-T1 (Netherlands), PIC (Spain), ASGC (Taiwan), RAL (UK) and BNL (USA), the Tier-2 facilities worldwide and large non-WLCG resource providers. Major contributors of computing resources are listed in Ref.~\cite{ATL-SOFT-PUB-2021-003}.
 

\printbibliography

\clearpage 
 
\begin{flushleft}
\hypersetup{urlcolor=black}
{\Large The ATLAS Collaboration}

\bigskip

\AtlasOrcid[0000-0002-6665-4934]{G.~Aad}$^\textrm{\scriptsize 98}$,    
\AtlasOrcid[0000-0002-5888-2734]{B.~Abbott}$^\textrm{\scriptsize 124}$,    
\AtlasOrcid[0000-0002-7248-3203]{D.C.~Abbott}$^\textrm{\scriptsize 99}$,    
\AtlasOrcid[0000-0002-2788-3822]{A.~Abed~Abud}$^\textrm{\scriptsize 34}$,    
\AtlasOrcid[0000-0002-1002-1652]{K.~Abeling}$^\textrm{\scriptsize 51}$,    
\AtlasOrcid[0000-0002-2987-4006]{D.K.~Abhayasinghe}$^\textrm{\scriptsize 91}$,    
\AtlasOrcid[0000-0002-8496-9294]{S.H.~Abidi}$^\textrm{\scriptsize 27}$,    
\AtlasOrcid[0000-0002-9987-2292]{A.~Aboulhorma}$^\textrm{\scriptsize 33e}$,    
\AtlasOrcid[0000-0001-5329-6640]{H.~Abramowicz}$^\textrm{\scriptsize 157}$,    
\AtlasOrcid[0000-0002-1599-2896]{H.~Abreu}$^\textrm{\scriptsize 156}$
% ...
\bigskip
\\

% ...
$^{27}$Physics Department, Brookhaven National Laboratory, Upton NY; United States of America.\\
% ...
$^{33}$$^{(a)}$Facult\'e des Sciences Ain Chock, R\'eseau Universitaire de Physique des Hautes Energies - Universit\'e Hassan II, Casablanca;$^{(b)}$Facult\'{e} des Sciences, Universit\'{e} Ibn-Tofail, K\'{e}nitra;$^{(c)}$Facult\'e des Sciences Semlalia, Universit\'e Cadi Ayyad, LPHEA-Marrakech;$^{(d)}$LPMR, Facult\'e des Sciences, Universit\'e Mohamed Premier, Oujda;$^{(e)}$Facult\'e des sciences, Universit\'e Mohammed V, Rabat;$^{(f)}$Mohammed VI Polytechnic University, Ben Guerir; Morocco.\\
$^{34}$CERN, Geneva; Switzerland.\\
% ...
$^{51}$II. Physikalisches Institut, Georg-August-Universit\"{a}t G\"ottingen, G\"ottingen; Germany.\\
% ...
$^{91}$Department of Physics, Royal Holloway University of London, Egham; United Kingdom.\\
% ...
$^{98}$CPPM, Aix-Marseille Universit\'e, CNRS/IN2P3, Marseille; France.\\
$^{99}$Department of Physics, University of Massachusetts, Amherst MA; United States of America.\\
% ...
$^{124}$Homer L. Dodge Department of Physics and Astronomy, University of Oklahoma, Norman OK; United States of America.\\
% ...
$^{156}$Department of Physics, Technion, Israel Institute of Technology, Haifa; Israel.\\
$^{157}$Raymond and Beverly Sackler School of Physics and Astronomy, Tel Aviv University, Tel Aviv; Israel.\\
% ...
$^{a}$ Also at Borough of Manhattan Community College, City University of New York, New York NY; United States of America.\\
$^{b}$ Also at Bruno Kessler Foundation, Trento; Italy.\\
$^{c}$ Also at Center for High Energy Physics, Peking University; China.\\
$^{d}$ Also at Centro Studi e Ricerche Enrico Fermi; Italy.\\
$^{e}$ Also at CERN, Geneva; Switzerland.\\
$^{f}$ Also at D\'epartement de Physique Nucl\'eaire et Corpusculaire, Universit\'e de Gen\`eve, Gen\`eve; Switzerland.\\
% ...

\end{flushleft}


\end{document}